\documentclass{article}

\usepackage[preprint]{neurips_2026}

\usepackage[utf8]{inputenc} 
\usepackage[T1]{fontenc}    
\usepackage{hyperref}       
\usepackage{url}            
\usepackage{booktabs}       
\usepackage{amsfonts}       
\usepackage{nicefrac}       
\usepackage{microtype}      
\usepackage{multirow}
\usepackage{listings}
\usepackage{caption}
\usepackage{booktabs}
\usepackage{pifont}
\usepackage{amssymb} 
\usepackage{xcolor}         
\usepackage{graphicx}
\usepackage{subcaption}
\usepackage{xcolor}
\usepackage{makecell}
\usepackage{amsmath} 
\usepackage[most]{tcolorbox}
\usepackage{listings}

\newtcblisting{promptbox}{
  enhanced,
  breakable,
  colback=gray!3,
  colframe=gray!45,
  boxrule=0.4pt,
  arc=2pt,
  left=5pt,
  right=5pt,
  top=5pt,
  bottom=5pt,
  listing only,
  listing options={
    basicstyle=\ttfamily\scriptsize,
    breaklines=true,
    columns=fullflexible,
    keepspaces=true,
    showstringspaces=false
  }
}
\title{ReasonBreak: Probing Vulnerabilities in Reasoning-Enabled Vision-Language-Action Models for Autonomous Driving}

\author{
  Mohammadreza Teymoorianfard\\
  University of Massachusetts Amherst \\
  \texttt{mteymoorianf@cs.umass.edu} \\
  \And Jean-Philippe Monteuuis\\
  Qualcomm\\
  \texttt{jmonteuu@qti.qualcomm.com}\\
  \And Jonathan Petit \\
  Qualcomm\\
  \texttt{petit@qti.qualcomm.com}\\
  \And Amir Houmansadr\\
  University of Massachusetts Amherst \\
  \texttt{amir@cs.umass.edu}\\
}

\begin{document}

\maketitle

\begin{abstract}

Vision-Language-Action (VLA) models with integrated reasoning have been proposed for end-to-end autonomous driving, assuming a tight coupling between reasoning and trajectory generation. However, the robustness of such systems under realistic input perturbations remains largely unexplored. We show that these models are highly vulnerable to realistic input perturbations, achieving up to \textbf{89\%} attack success rate (ASR) on reasoning and up to \textbf{72\%} on trajectory manipulation in closed-loop simulation, leading to increased collision rates and degraded safety metrics. Using NVIDIA's recent Alpamayo models as representative industry-developed VLAs, we conduct the first systematic black-box study of reasoning-enabled VLA models under realistic textual input corruptions, evaluating their impact on reasoning and driving behavior. We introduce a reasoning-aware evaluation framework capturing both semantic and structural aspects of reasoning, along with safety-centric measures. We also introduce a benchmark for evaluating attacks and defenses on reasoning–trajectory interactions in autonomous driving. Our results highlight the need for rigorous evaluation and improved defenses to ensure the safety of reasoning-enabled VLA systems in autonomous driving. 

\end{abstract}

\begin{center}
    \includegraphics[width=0.88\linewidth]{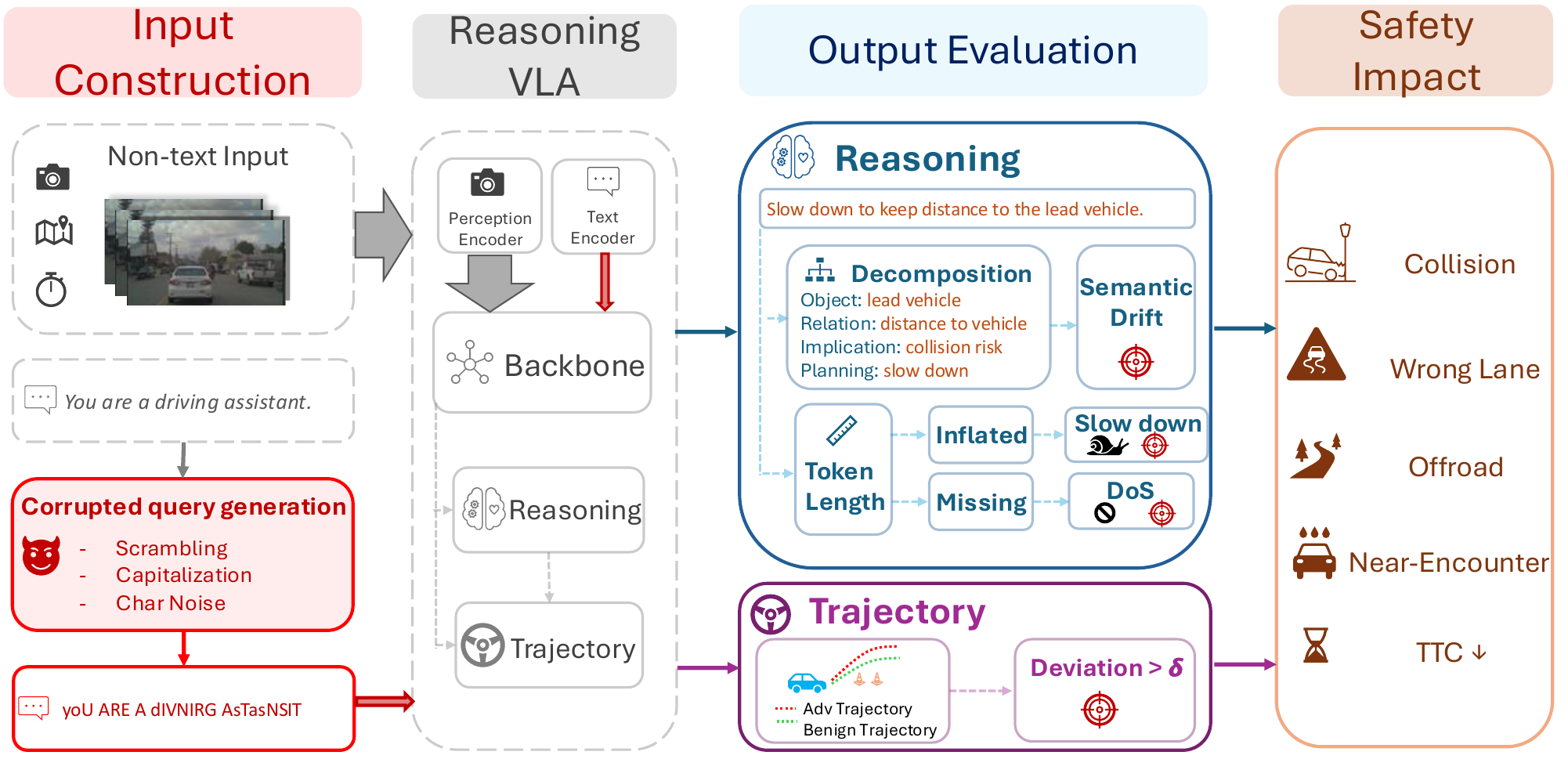}
    \captionof{figure}{
ReasonBreak pipeline. We corrupt only the small textual input channel while keeping the dominant non-text fixed, then measure reasoning shifts, trajectory degradation, and safety impact.
}
    \label{fig:pipeline}
\end{center}

\section{Introduction}

Vision-Language-Action (VLA) models are emerging as a prominent direction for end-to-end autonomous driving. This trend is increasingly visible in industry-scale systems: NVIDIA's Alpamayo family introduces reasoning-enabled VLA models that couple grounded reasoning with trajectory prediction~\cite{wang2025alpamayo, nvidia_alpamayo_announcement_2026}. By incorporating language, VLA systems augment conventional end-to-end driving pipelines~\cite{bojarski2016end, codevilla2019exploring, dosovitskiy2017carla, codevilla2018end, chitta2022transfuser} with high-level context and task constraints~\cite{kim2024openvla, black2024pi_0, qu2025spatialvla, zhou2025autovla, wang2025alpamayo}; recent variants further expose intermediate reasoning before action generation~\cite{deng2025graspvla, lee2025molmoact, yang2025instructvla, wang2025alpamayo}.

The integration of reasoning into VLA-based driving systems introduces a new and largely unexplored dimension of risk, as intermediate reasoning is tightly coupled with downstream trajectory generation. However, existing studies on VLA vulnerabilities have primarily focused on robotic manipulation settings and models without explicit reasoning components \cite{wang2025freezevla, zhou2025badvla, lu2026phantom, lu2025robots, zhang2025attention, jones2025adversarial}. These settings differ fundamentally from autonomous driving in both objectives and failure modes. In driving systems, errors directly correspond to safety violations \cite{ross2011reduction, codevilla2019exploring} rather than task-level inaccuracies. As a result, the role of reasoning as a potential source of vulnerability in safety-critical autonomous driving systems remains largely unexplored.

In real-world deployment, VLA-based driving systems may receive textual inputs through imperfect language interfaces, including speech-to-text transcription, in-vehicle dialogue systems, and downstream text preprocessing pipelines~\cite{anjum2023improving, dai2022ci, weng2016conversational}. Such interfaces do not provide a clean, canonical text channel: environmental noise, speaker variability, accent differences, segmentation errors, and normalization artifacts can all change the surface form of the input text command ~\cite{smith2017speech, shah2024speech, del2023accents, southwell2022challenges, cui2021approach, chai2023comparison}.
Consequently, VLA models such as Alpamayo, which operate on textual instructions, are naturally exposed to noisy and imperfect input text in realistic settings.

Prior work has shown that autoregressive language models are sensitive to surface-form perturbations, where small changes to spelling, casing, or character composition can significantly alter model behavior, particularly in reasoning tasks~\cite{gan2024reasoning, alahmari2025large, liu2025evaluating}. Motivated by this observation, we study VLA vulnerabilities under a query-based black-box setting that reflects these practical constraints, and ask whether realistic input text perturbations can influence reasoning and lead to unsafe behavior.

We find that corrupted inputs disrupt the intended reasoning--action coupling in reasoning-enabled driving VLAs: reasoning shifts and trajectory deviations are only weakly correlated, with manipulated reasoning often failing to induce corresponding control changes and large trajectory deviations sometimes occurring with little reasoning change. This exposes reasoning as a practically exploitable failure surface. Under closed-loop interaction, realistic textual corruptions achieve up to 72\% ASR for trajectory manipulation and up to 62\% ASR for semantic reasoning manipulation, while also inducing structural reasoning failures such as length inflation and denial-of-service (DoS), where model refuses to generate any token. These perturbations translate into safety-critical degradation, increasing downstream failures such as collisions, off-road behavior, and wrong-lane events.

We make the following contributions:

\begin{itemize}

\item \textbf{First systematic study of reasoning vulnerabilities in VLA-based autonomous driving.}
We study reasoning-enabled VLA models in safety-critical autonomous driving using a safety-centric evaluation pipeline that combines fine-grained driving metrics (collision rate, near-encounter, time-to-collision) with reasoning-aware analysis. We find that (i) both reasoning and trajectory can be manipulated to induce unsafe behavior, (ii) reasoning and trajectory are weakly correlated under manipulated inputs despite architectural coupling, and (iii) RL post-training improves benign robustness but can amplify safety risks when attacks succeed. We further show that lightweight rule-based filtering substantially mitigates these vulnerabilities.

\item \textbf{Reasoning as an attack surface under realistic black-box constraints.}
We identify reasoning as a distinct and previously underexplored attack surface in VLA systems, and show that both semantic and structural properties of reasoning (e.g., content and length) can be manipulated under realistic input perturbations. We study this in a constrained, query-based black-box setting with limited input control, and adapt reasoning attacks (e.g., slowdown and DoS) to demonstrate their impact on safety-critical behavior.

\item \textbf{Benchmark and dataset for reasoning–safety evaluation in autonomous driving.}
We release the first benchmark and dataset for evaluating reasoning–safety interactions in VLA-based autonomous driving, mapping input conditions to intermediate reasoning outputs and safety-critical outcomes.

\end{itemize}

\section{Background and Related Work}

\subsection{Vulnerabilities of VLAs}

Vision-Language-Action (VLA) models jointly process visual observations and language inputs to generate sequential actions, enabling end-to-end decision making in embodied settings~\cite{kim2024openvla,black2024pi_0}. Recent work has explored such models in robotic manipulation~\cite{kim2024openvla,qu2025spatialvla,black2024pi_0,wen2025tinyvla,liu2025hybridvla} and autonomous driving~\cite{zhou2025autovla,zhou2026opendrivevla,rowe2025poutine,wang2026hist,wang2025alpamayo}. In this work, we focus on Alpamayo~\cite{wang2025alpamayo}, a recent reasoning-enabled VLA system for autonomous driving.

Given their deployment in physical systems, VLA models expose multiple safety-critical attack surfaces that have recently been explored. Prior work has identified vulnerabilities at the perception level, including physical sensor attacks \cite{lu2026phantom} and adversarial patches \cite{lu2025robots, zhang2025attention, wang2025exploring, xu2025model}, as well as at the model level through action-freezing \cite{wang2025freezevla} and backdoor attacks \cite{zhou2025badvla}. Language-based attacks can also induce unsafe behaviors \cite{jones2025adversarial}, alongside agentic black-box frameworks that exploit model interactions \cite{wu2026saber}. More recently, multimodal attacks that jointly manipulate visual and textual inputs have been shown to amplify attack effectiveness and bypass alignment mechanisms \cite{yan2025alignment}.

However, existing studies on VLA vulnerabilities have primarily focused on robotic manipulation settings and models without explicit reasoning components \cite{wang2025exploring, yan2025alignment, zhang2025attention}. These works typically target input-level or policy-level weaknesses and evaluate failures in terms of task-level degradation or trajectory deviation, where small errors do not necessarily imply unsafe behavior. In contrast, autonomous driving requires evaluating safety-critical outcomes, where even minor perturbations can lead to severe consequences \cite{ross2011reduction, codevilla2019exploring}. Recent work has examined robustness in autonomous-driving VLMs and language-driven driving systems, including black-box/visual attacks, decision-making robustness, and instruction counterfactual robustness~\cite{wang2025black,zhang2024visual,cheng2024decictor,hamid2026icr}. However, these studies do not address reasoning-enabled driving VLAs as dual-output systems, where textual corruptions may affect both intermediate reasoning and trajectory behavior. In this work, we study this gap.

\subsection{Reasoning Vulnerabilities}

With the emergence of reasoning capabilities in large language and vision-language models, recent work has begun to explore vulnerabilities introduced by this new attack surface \cite{si2025excessive}. Prior studies show that reasoning processes can be manipulated in multiple ways, such as inducing excessively long reasoning chains to increase latency (slowdown attacks) \cite{kumar2025overthink, liu2026reasoningbomb} or disrupting reasoning through premature termination (denial-of-service attacks) \cite{cui2025practical}. These findings highlight that reasoning itself can be directly targeted to influence model behavior.

However, these attacks are primarily studied in text-only settings, where reasoning is represented as free-form natural language. In contrast, reasoning in VLA models is grounded in structured, task-relevant representations and tightly integrated with downstream action generation \cite{wang2025alpamayo, nvidia_alpamayo_announcement_2026}. Moreover, unlike LLMs and VLMs, reasoning in VLAs is not directly controllable through prompts, making it difficult to apply existing reasoning attacks in a straightforward manner.

To date, reasoning as an attack surface in VLA models remains largely underexplored. TRAP \cite{huang2026trap} provides an initial step by showing that manipulating intermediate reasoning can influence downstream actions, but focuses on robotic manipulation settings and relies on perception-driven, white-box attacks. In contrast, we study reasoning-level vulnerabilities in VLA systems under a realistic black-box setting, with a focus on safety-critical autonomous driving.

\section{Threat Model}

\begin{figure}
    \centering
    \includegraphics[width=0.78\linewidth]{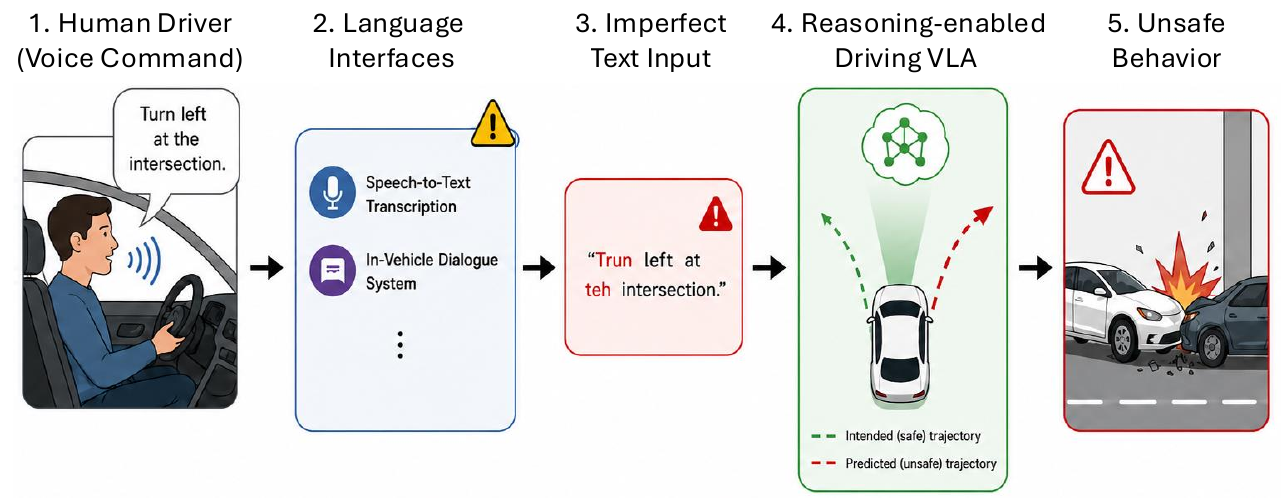}
    \caption{Threat model: imperfect language interfaces can expose reasoning-enabled driving models to malformed textual inputs, leading to unsafe behavior.}
    \label{fig:threat_model}
\end{figure}

Figure~\ref{fig:threat_model} illustrates our threat model: imperfect language interfaces can produce malformed text inputs that perturb outputs and lead to unsafe behavior.

\textbf{Adversary Goal:}
The adversary aims to induce deviations in the model’s behavior by manipulating its reasoning or trajectory outputs. Specifically, the goal is to influence reasoning (e.g., its semantic content or structure) or the predicted trajectory in a way that leads to degraded performance or unsafe outcomes.

\textbf{Adversary Capabilities:}
We assume a query-based black-box setting, where the adversary has no access to model parameters, architecture, internal states, logits, or confidence scores. The adversary can only interact with the model through input queries and observe its outputs, including generated reasoning and predicted trajectories.

\textbf{Assumptions:}
The adversary acts on the digital textual input received by the model, not on physical audio or visual sensors. Its control is restricted to semantically preserving corruptions: it cannot inject arbitrary malicious instructions or nonsensical tokens, and can only modify the surface form of an input through capitalization changes, word scrambling, or character-level noise.

\section{Methodology}

Figure~\ref{fig:pipeline} summarizes our pipeline: textual corruptions are applied while non-text inputs are fixed, and the reasoning and trajectory outputs are evaluated for attack success and downstream safety impact.

\subsection{Problem Setup}

We consider a reasoning-enabled VLA model \(f\) that maps multimodal inputs to both intermediate reasoning and trajectory outputs~\cite{wang2025alpamayo}. Given visual observations \(v\) and textual input \(t\), the model produces reasoning \(r\) and trajectory \(\tau\): \((r,\tau)=f(v,t)\). We study textual perturbations \(t' \in \mathcal{T}(t)\) that preserve the intended meaning of \(t\) while introducing small realistic variations, such as character-level noise or word-level modifications, keeping \(v\) fixed. The perturbed output is \((r',\tau')=f(v,t')\), and our goal is to measure how such perturbations alter reasoning and trajectory behavior, and whether they induce safety-critical degradation.

\subsection{Query-Based Perturbation Pipeline}

\paragraph{Open-Loop (Capability Upper Bound).}
In the open-loop setting, we evaluate the extent to which model behavior can be manipulated under a query-based exploration of the perturbation space. Specifically, we adopt a Best-of-\(N\) strategy~\cite{hughesbest}, where a sequence of perturbed inputs \( \{t_i\}_{i=1}^{N} \), with \( t_i \sim \mathcal{T}(t) \), are evaluated as:
\[
(r_i, \tau_i) = f(v, t_i).
\]
The search terminates early if a predefined objective is satisfied. To obtain stable estimates, we repeat this process and aggregate results using bootstrap-style resampling~\cite{efron1992bootstrap}.

This setting provides a controlled probe of model vulnerability, estimating the upper bound on attack success under constrained but adaptive perturbations.

\paragraph{Closed-Loop (Deployment Realism).}
In the closed-loop setting, we evaluate perturbations under a realistic deployment regime without query-based search. At each simulation step \(k\), we sample a perturbed textual input \(t'_k \sim \mathcal{T}(t_k)\) and query the model once:
\[
(r_k, \tau_k) = f(v_k, t'_k).
\]
The predicted trajectory is then applied in the simulator, influencing future observations \(v_{k+1}\). Thus, unlike open-loop evaluation, closed-loop evaluation does not search over multiple perturbations at each step; it uses a single randomly corrupted input per model invocation. This design reflects real-time constraints, where the model is invoked at high frequency and multi-query exploration at each step would be unrealistic. It captures whether simple perturbations, without optimization or selection, can accumulate over time and lead to safety-critical outcomes under continuous interaction.

Together, these two settings provide a complementary view: open-loop evaluation characterizes the manipulability of the model under exploration, while closed-loop evaluation measures the realized impact of perturbations under realistic deployment dynamics.

\subsection{Objectives \& Safety Impact}
We evaluate textual perturbations on two output surfaces of reasoning-enabled VLA models: \emph{reasoning} and \emph{trajectory}. For reasoning, we measure semantic shifts across four task-relevant fields---object, relation, implication, and planning---as well as structural failures such as reasoning-length inflation and missing reasoning outputs. For trajectory, we measure perturbation-induced degradation relative to the model's benign behavior on the same scenario, using ADE in open-loop evaluation and realized ego-position deviation in closed-loop evaluation. The precise reference definitions, thresholds, and attack success criteria are provided in Section~\ref{sec:asr_metrics}.

Safety is not an explicit attack objective. Instead, after identifying successful reasoning or trajectory manipulations, we quantify their downstream safety impact. In open-loop evaluation, we report collision rate, near-encounter rate, and minimum time-to-collision (min-TTC); in closed-loop evaluation, we report scenario-level AlpaSim metrics~\cite{alpasim_2025}, including collision, off-road, and wrong-lane rate.

\section{Evaluation Protocol and Success Criteria}
\label{sec:asr_metrics}

Existing VLA robustness studies largely target task-level failures in robotic manipulation, often without explicit reasoning outputs~\cite{wang2025freezevla, zhou2025badvla, lu2026phantom, lu2025robots, zhang2025attention, jones2025adversarial}. In contrast, reasoning-enabled autonomous-driving VLAs expose two coupled but distinct surfaces: reasoning and trajectory. We therefore introduce an evaluation protocol that defines attack success over semantic reasoning shifts, structural reasoning failures, trajectory degradation, and closed-loop rollout failures. Throughout this section, \(\mathcal{C}\) denotes a binary per-target success condition, and \(\mathcal{A}_{\mathrm{cl}}\) denotes rollout-level closed-loop success.

\subsection{Output-Surface Success Conditions}

\paragraph{Reference.}

For each scenario, let \(v\) be the visual input, \(t\) the clean text, and \(t'\) its perturbed counterpart. Let \((r^0,\tau^0)\) and \((r',\tau')\) denote the model outputs under clean and perturbed text, respectively:
\[
(r^0,\tau^0)=f(v,t), \qquad (r',\tau')=f(v,t').
\]
We use \((r^0,\tau^0)\) as the benign reference.

\paragraph{Reasoning Evaluation.}
We evaluate semantic and structural changes in reasoning. For semantic evaluation, an LLM-based evaluator decomposes each reasoning output into four task-relevant fields: \emph{object}, \emph{relation}, \emph{implication}, and \emph{planning}, capturing the key entity affecting the driving decision, its relation to the ego vehicle, the implied risk or constraint, and the planned action. For example, ``Slow down to keep distance to the lead vehicle'' is decomposed as: \emph{object}: lead vehicle; \emph{relation}: distance to vehicle; \emph{implication}: collision risk; and \emph{planning}: slow down.

For each target field \(c\), the evaluator assigns a semantic deviation score \(S_{\text{sem}}^{c}(r',r^0)\in[0,1]\), where larger values indicate greater semantic change from the benign reasoning. For structural evaluation, we consider \emph{slowdown}, where \(r'\) is substantially longer than \(r^0\), and \emph{DoS}, where the model produces no reasoning tokens.

For a reasoning target \(c\), we instantiate the binary success condition as:
\[
\mathcal{C}^{\text{Reason}}_{c}(r', r^0) =
\begin{cases}
\mathbf{1}\!\left(S_{\text{sem}}^{c}(r', r^0) > \delta_{\text{sem}}\right), 
& c \in \{\text{obj},\text{rel},\text{impl},\text{plan},\text{all}\},\\[4pt]
\mathbf{1}\!\left(\frac{|r'|}{|r^0|} > \rho\right), 
& c = \text{slowdown},\\[4pt]
\mathbf{1}\!\left(|r'|=0\right), 
& c = \text{DoS}.
\end{cases}
\]
Here, \(\delta_{\text{sem}}\) is the semantic-deviation threshold, \(\rho\) is the slowdown 
threshold, and \(|r|\) denotes the number of generated reasoning tokens.

\paragraph{Trajectory Evaluation.}
We evaluate trajectory degradation with a setting-specific error \(d_{\mathrm{traj}}(\cdot,\cdot)\) computed against the ground truth. In open-loop evaluation, \(d_{\mathrm{traj}}\) is ADE between the predicted and ground-truth trajectories, following prior work~\cite{wang2025alpamayo}; in closed-loop evaluation, it is the distance between the realized ego position and the corresponding ground-truth position at each time step. 

Let \(o^{0}\) denote the benign trajectory output or ego state, \(o'\) the corresponding output under perturbation, and \(o^{\mathrm{gt}}\) the ground truth. We define the perturbation-induced excess trajectory error as
\[
e_{\mathrm{traj}} = d_{\mathrm{traj}}(o', o^{\mathrm{gt}}) - d_{\mathrm{traj}}(o^{0}, o^{\mathrm{gt}}).
\]
The trajectory success condition is
\[
\mathcal{C}^{\mathrm{Traj}}_{\mathrm{dev}} = \mathbf{1}\!\left(e_{\mathrm{traj}} > \delta_{\mathrm{traj}}\right),
\]
where \(\delta_{\mathrm{traj}}\) is the trajectory-degradation threshold. Thus, a trajectory attack succeeds only when the perturbation causes additional error beyond the model's nominal behavior.

\subsection{Attack Success Rate (ASR)}
ASR is defined differently for open- and closed-loop settings.

\textbf{Open-loop.}
Given \(N\) queried perturbations \(\{t_i\}_{i=1}^{N}\), an attack succeeds if any query satisfies the target condition:
\[
\text{ASR}_{\text{open}}^{(s,c)}
=
\mathbb{E}\left[
\mathbf{1}\left(
\exists i \leq N :
\mathcal{C}^{(s,c)}_i = 1
\right)
\right],
\]
where \(s\in\{\mathrm{Reason},\mathrm{Traj}\}\), \(c\) denotes the target objective, and \(\mathcal{C}^{(s,c)}_i\) is the target-specific success condition evaluated on the \(i\)-th perturbed output.

\textbf{Closed-loop.}
In closed-loop evaluation, each rollout produces a time sequence of outputs, and even benign rollouts can deviate from the reference due to simulator stochasticity and rollout instability. We therefore measure success using \emph{excess deviation}: the deviation under perturbation minus the deviation already present in the benign rollout.

Let \(\{o_t^0\}_{t=1}^{T}\) and \(\{o_t'\}_{t=1}^{T}\) denote the benign and perturbed output sequences, where \(o_t\) may be reasoning \(r_t\) or trajectory output \(\tau_t\). For surface \(s\) and target \(c\), let \(D_{\mathrm{cl}}^{(s,c)}(\cdot,\cdot)\) be the closed-loop target-specific deviation function. We define the per-step excess deviation as
\[
\Delta_t^{(s,c)}
=
D_{\mathrm{cl}}^{(s,c)}(o_t', o_t^{\mathrm{ref}})
-
D_{\mathrm{cl}}^{(s,c)}(o_t^0, o_t^{\mathrm{ref}}),
\]
where \(o_t^{\mathrm{ref}}\) is the aligned reference at time \(t\). This subtraction prevents nominal closed-loop drift or simulator variability from being counted as attack success.

Due to closed-loop compounding, small early deviations can accumulate into later failures, so rollout-averaged deviation can understate attack success; conversely, relying on a single time step can overstate success due to transient noise. We therefore define a rollout-level success indicator \(\mathcal{A}_{\mathrm{cl}}^{(s,c)}\). A closed-loop attack is successful if there exists a contiguous window \(W\) of at least \(w\) steps for which the average excess deviation exceeds the target-specific threshold:
\[
\mathcal{A}_{\mathrm{cl}}^{(s,c)}
=
\mathbf{1}\left(
\exists W \subseteq \{1,\dots,T\},\ |W|\geq w :
\frac{1}{|W|}
\sum_{t\in W}
\Delta_t^{(s,c)}
>
\delta_{\mathrm{cl}}^{(s,c)}
\right).
\]
The closed-loop ASR is then
\[
\text{ASR}_{\mathrm{cl}}^{(s,c)}
=
\mathbb{E}\left[
\mathcal{A}_{\mathrm{cl}}^{(s,c)}
\right].
\]

\section{Experiments}

\subsection{Experimental Setup} \label{sec:exp_setup}

\subsubsection{Models}
We evaluate Alpamayo1 and Alpamayo1.5, NVIDIA's recent reasoning-enabled autonomous-driving VLAs~\cite{wang2025alpamayo,nvidia_alpamayo_announcement_2026}. Alpamayo1 is trained with supervised learning, while Alpamayo1.5 adds RL post-training. These models fit our study because they expose explicit, structured reasoning rather than free-form text, tightly couple reasoning with trajectory generation, and provide public datasets and simulation tooling. To our knowledge, no other publicly available industry-developed driving VLA currently offers this combination of structured reasoning outputs, trajectory prediction, and reproducible open-/closed-loop evaluation support.

\subsubsection{Dataset \& Simulator}

For open-loop evaluation, we use 195 samples from the NVIDIA Physical AI Autonomous Vehicles dataset~\cite{nvidia_physicalai_av_2025}, which provides paired visual/textual inputs, reference trajectories, and temporally annotated object bounding boxes. Since the released dataset does not provide reasoning annotations or precomputed safety-critical metrics, we use each model's default, unperturbed output as the benign reference for reasoning comparisons and derive collision, near-encounter, and time-to-collision metrics from the bounding boxes. For closed-loop evaluation, we use 50 clips from the NVIDIA Physical AI Autonomous Vehicles NuRec dataset~\cite{nvidia_physicalai_nurec_2025} and deploy them in AlpaSim~\cite{alpasim_2025}. 

\subsection{Open-Loop Analysis}
\label{sec:open-loop}

We first analyze model behavior under open-loop conditions. Table~\ref{tab:open-loop_main} reports the ASR, the average number of queries to success. Table \ref{tab:open-loop_main-apdx} safety metrics as well. ASR is computed over the full dataset, while safety metrics are reported as absolute values with changes ($\Delta$) measured relative to outputs from the same model on uncorrupted inputs for the successful attack instances.

\paragraph{Reasoning and Trajectory Manipulation.}
Both models exhibit substantial vulnerability to realistic textual perturbations. For Alpamayo1, semantic reasoning attacks achieve high ASRs across object, relation, implication, and planning targets (0.76--0.89), showing that small perturbations can reliably alter the generated reasoning. Alpamayo1.5 is more robust, with lower but still non-trivial ASRs (0.42--0.63). Structural reasoning attacks are harder, but still succeed in a subset of cases: slowdown occurs more frequently than DoS, which remains rare but observable. Direct trajectory attacks also achieve meaningful success rates and consistently degrade driving behavior, confirming trajectory prediction as the most direct attack surface.

\paragraph{Safety Impact.}
Reasoning manipulation has model-dependent safety effects. For Alpamayo1, successful reasoning attacks often produce inconsistent changes in safety metrics, with both positive and negative $\Delta$ values, suggesting that altered reasoning does not reliably translate into degraded control. In contrast, Alpamayo1.5 shows more consistent safety degradation under successful reasoning attacks, including increased collision rates and reduced minimum TTC across several targets. This suggests that its RL post-training may improve robustness overall, but also be more susceptible to safety-critical failures once this coupling is disrupted. Structural attacks, although less frequent, tend to induce severe degradation when successful, especially for Alpamayo1.5. By comparison, trajectory attacks lead to more stable and pronounced safety degradation across both models, increasing collision and near-encounter rates while reducing minimum TTC.

\paragraph{Reasoning--Trajectory Coupling.}
Figure~\ref{fig:correlation} and \ref{fig:correlation-a15} further examine the relationship between reasoning manipulation and trajectory deviation under corrupted inputs. Although both surfaces can be manipulated, their correlation remains weak: changes in reasoning do not consistently induce large trajectory deviations, and large trajectory deviations do not necessarily coincide with large reasoning shifts. At the same time, reasoning components remain strongly correlated with each other, indicating that the generated reasoning is internally coherent. This suggests a weak translation from reasoning to control under perturbation, which contrasts with the intended role of reasoning in reasoning-enabled VLA systems. Importantly, weak correlation does not imply low risk: even small trajectory deviations associated with reasoning changes can still compound into safety-critical outcomes.

\paragraph{Latency Correlation.}
Figure~\ref{fig:latency} shows the strong linear correlation between the number of generated reasoning tokens and latency.
This shows that slowdown attacks not only affect the reasoning process but also introduce additional latency overhead in real-time settings such as autonomous driving.

\begin{table}[t]
\centering
\caption{Open-loop ASR and query efficiency across reasoning and trajectory targets.}
\label{tab:open-loop_main}
\small
\setlength{\tabcolsep}{5pt}
\resizebox{0.6\textwidth}{!}{%
\begin{tabular}{@{}l l l l c c@{}}
\toprule
\multirow{2}{*}{\textbf{Model}} 
& \multirow{2}{*}{\textbf{Surface}} 
& \multirow{2}{*}{\textbf{Objective}} 
& \multirow{2}{*}{\textbf{Target}}
& \multirow{2}{*}{\textbf{ASR} $\uparrow$}
& \multirow{2}{*}{\textbf{Queries} $\downarrow$} \\
&&&&& \\
\midrule

\multirow{9}{*}{Alp1}
& \multirow{7}{*}{Reasoning}
& \multirow{5}{*}{Semantic}
& Object      & 0.765 & 10.2 \\
& & & Relation    & 0.889 &  6.8 \\
& & & Implication & 0.850 &  7.6 \\
& & & Planning    & 0.832 &  8.3 \\
& & & Overall     & 0.836 &  9.4 \\
\cmidrule(l){3-6}
& & \multirow{2}{*}{Structural}
& Slowdown & 0.248 & 31.9 \\
& & & DoS      & 0.047 & 37.4 \\
\cmidrule(l){2-6}
& Trajectory
& Deviation
& ADE & 0.336 & 18.6 \\

\midrule

\multirow{9}{*}{Alp1.5}
& \multirow{7}{*}{Reasoning}
& \multirow{5}{*}{Semantic}
& Object      & 0.436 & 10.5 \\
& & & Relation    & 0.626 &  9.8 \\
& & & Implication & 0.520 &  9.8 \\
& & & Planning    & 0.429 & 10.5 \\
& & & Overall     & 0.422 &  8.9 \\
\cmidrule(l){3-6}
& & \multirow{2}{*}{Structural}
& Slowdown & 0.102 & 12.1 \\
& & & DoS      & 0.088 & 15.8 \\
\cmidrule(l){2-6}
& Trajectory
& Deviation
& ADE & 0.115 & 21.5 \\

\bottomrule
\end{tabular}
}
\end{table}

\subsection{Closed-Loop Analysis} \label{sec:closed-loop}

\begin{figure}[t]
\centering

\begin{minipage}{0.42\linewidth}
\centering
\captionof{table}{Closed-loop attack success rates (ASR) across trajectory and reasoning targets.}
\label{tab:closed-loop-main}
\vspace{2pt}

\small
\setlength{\tabcolsep}{3pt}
\renewcommand{\arraystretch}{1.1}

\resizebox{\linewidth}{!}{%
\begin{tabular}{ccccccccc}
\toprule
\multirow{4}{*}{\textbf{Model}} 
& \multirow{3}{*}{\textbf{Traj}}
& \multicolumn{7}{c}{\textbf{Reasoning}} \\
\cmidrule(lr){3-9}

& & \multicolumn{5}{c}{Semantic}
& \multicolumn{2}{c}{Structural} \\
\cmidrule(lr){2-2} \cmidrule(lr){3-7} \cmidrule(lr){8-9}

& Dev & Obj & Rel & Impl & Plan & All & Slow & DoS \\
\midrule
Alp1   & 0.72 & 0.48 & 0.58 & 0.54 & 0.62 & 0.56 & 0.08 & 0.0 \\
Alp1.5 & 0.48 & 0.42 & 0.44 & 0.40 & 0.42 & 0.40 & 0.20 & 0.0 \\
\bottomrule
\end{tabular}
}
\end{minipage}
\hfill
\begin{minipage}{0.31\linewidth}
\centering

\includegraphics[width=\linewidth]{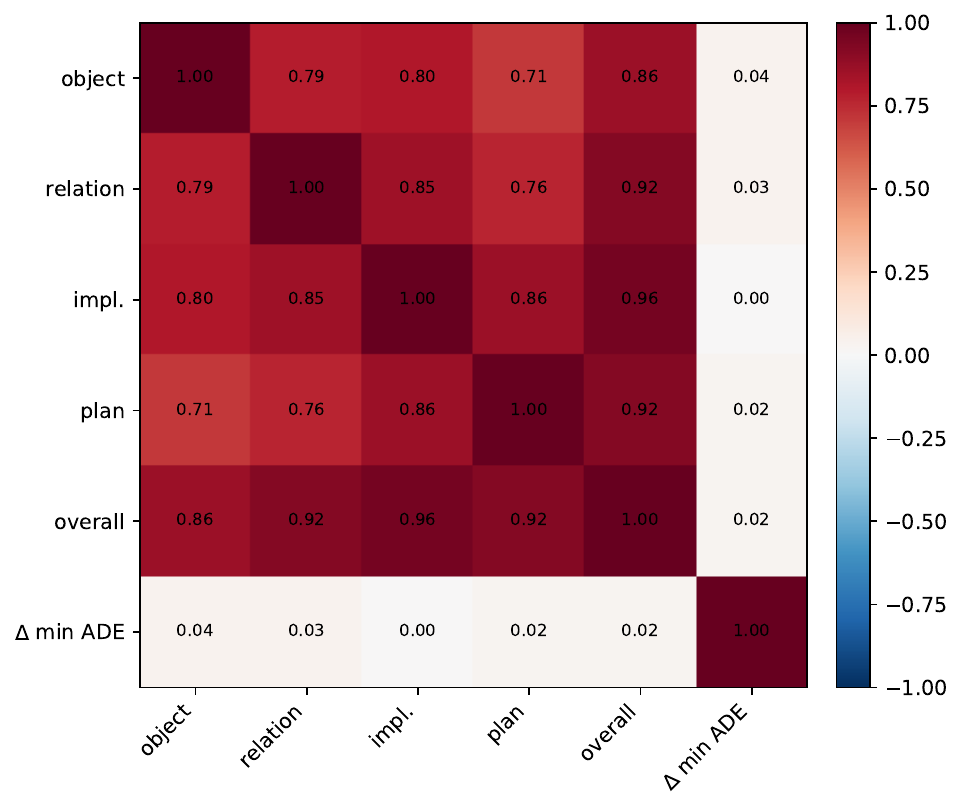}
\caption{Alp1: Reasoning-trajectory correlation}
\label{fig:correlation}

\end{minipage}

\end{figure}

\paragraph{Attack Effectiveness.}
Table~\ref{tab:closed-loop-main} reports ASR across trajectory and reasoning targets, in the closed-loop setting. Trajectory manipulation is the most effective closed-loop attack surface, achieving the highest ASR for both Alpamayo1 (72.0\%) and Alpamayo1.5 (48.0\%). Reasoning manipulation remains feasible, with semantic attacks reaching 48.0\%--62.0\% ASR for Alpamayo1 and 40.0\%--44.0\% for Alpamayo1.5, indicating reduced but persistent vulnerability under closed-loop interaction. Structural attacks are less effective: slowdown succeeds at modest rates, while DoS is not observed, consistent with the open-loop finding that fully suppressing reasoning is rare. The higher slowdown ASR in Alpamayo1.5 suggests that RL post-training may improve semantic robustness while increasing sensitivity to structural perturbations.

\paragraph{Safety Impact.}
We further analyze the downstream safety impact of successful closed-loop attacks in Appendix~\ref{apdx:closed_safety}. Figures~\ref{fig:closed_traj_safety} and~\ref{fig:closed_reas_safety} show that successful perturbations often destabilize model behavior across incident types, typically introducing new safety violations. When the benign rollout is already unsafe, attacks also tend to accelerate failure onset: Table~\ref{tab:closed_timing} shows that wrong-lane and off-road events occur earlier under perturbation for both models. Finally, Table~\ref{tab:closed_incident_window} shows that different manipulated surfaces correlate with different incident types near failure onset; trajectory deviations are more aligned with some events, while reasoning shifts are more prominent for others. This suggests that closed-loop safety degradation is not driven by a single failure channel, but by surface- and incident-dependent interactions between reasoning, trajectory, and rollout dynamics.

\subsection{More Analysis and Ablations}

 We evaluate the effect of input variations, including semantic changes, augmentation strength, and alternative input channels. With the semantically malicious prompts without textual corruption (Appendix~\ref{apdx:mal_vs_neutral}), we observe that both models remain relatively robust to such inputs, largely ignoring malicious intent when the text is clean. Also, when augmentation is being applied to such inputs across different semantic prompt variations, we do not observe considerable changes in ASR or safety metrics compared to the semantically neutral input.

As shown in Appendix~\ref{apdx:aug} and Appendix~\ref{apdx:input_token_length}, stronger augmentations and longer inputs consistently increase ASR, suggesting that both noise level and input surface area amplify attack effectiveness. Alpamayo1.5 additionally accepts navigation (Nav) commands, which provide another pathway for perturbation. As shown in Appendix~\ref{apdx:nav}, Nav-based attacks achieve comparable ASR and induce similar safety degradation, indicating that auxiliary inputs form an additional input attack surface.

\section{Mitigation}
\begin{figure}[t]
\centering

\begin{minipage}{0.3\linewidth}
\centering
\small
\setlength{\tabcolsep}{4pt}
\renewcommand{\arraystretch}{1.1}
\captionof{table}{ASR under baseline ($\times$) and defense ($\checkmark$).}

\begin{tabular}{ccccccccc}
\toprule
\multirow{4}{*}{\textbf{Def}} 
& \multirow{3}{*}{\textbf{Traj}}
& \multicolumn{7}{c}{\textbf{Reasoning}} \\

\cmidrule(lr){3-9}

& & \multicolumn{5}{c}{Semantic}
& \multicolumn{2}{c}{Structural} \\

\cmidrule(lr){2-2} \cmidrule(lr){3-7} \cmidrule(lr){8-9}

& ADE & Obj & Rel & Impl & Plan & All
& Slow & DoS \\
\midrule

\ding{55} 
& 0.336
& 0.765 & 0.889 & 0.850 & 0.832 & 0.836
& 0.248 & 0.047 \\

\ding{51} 
& 0.142
& 0.425 & 0.610 & 0.426 & 0.390 & 0.428
& 0.038 & 0.004 \\

\bottomrule
\end{tabular}

\label{tab:defence}

\end{minipage}
\hfill
\begin{minipage}{0.22\linewidth}
\centering
\includegraphics[width=\linewidth]{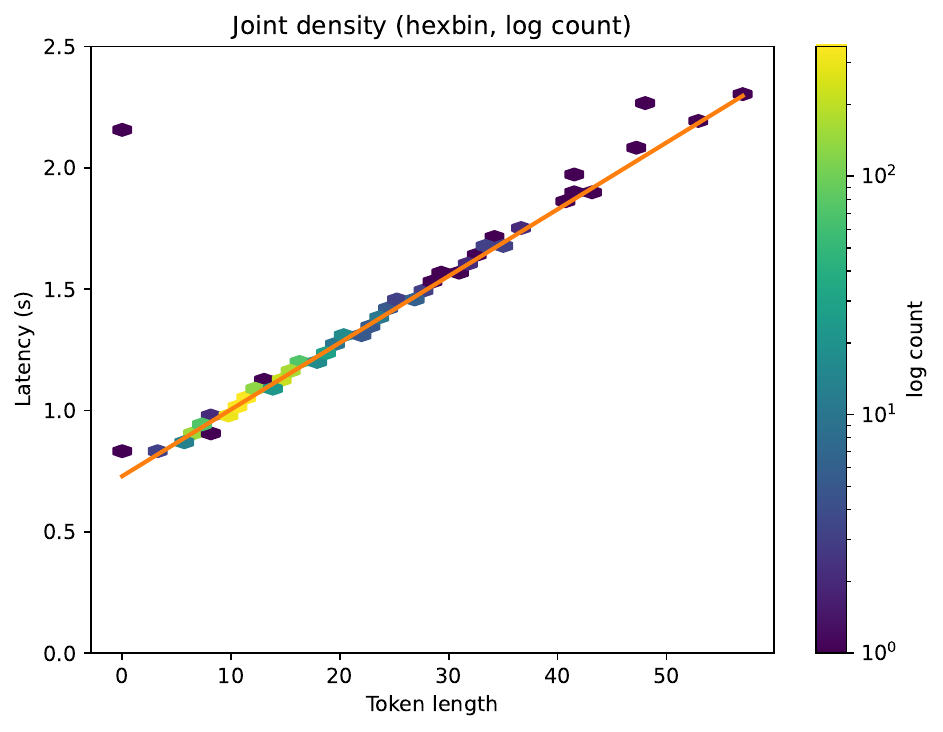}
\caption{Inference latency vs.\ reasoning length on Alp1.}
\label{fig:latency}
\end{minipage}

\end{figure}

Despite growing work on VLA attacks, defenses for these systems remain largely unexplored, especially under realistic textual corruptions. We introduce a lightweight input-normalization defense for this setting. Since our attacks preserve intent but corrupt the text surface, a natural defense is to canonicalize the input before it reaches the model. LLM-based correction is undesirable in real-time driving due to latency, while perplexity-based filtering requires threshold selection and discards inputs rather than correcting them~\cite{jones2025adversarial}. We therefore use a deterministic rule-based normalizer: given \(t\), we construct \(\tilde{t}\) by lowercasing, removing non-alphanumeric characters, normalizing whitespace, applying token-level fuzzy correction with a small closed vocabulary, and restoring minimal formatting. This approach is deterministic, efficient, and introduces almost no additional latency. As shown in Table~\ref{tab:defence}, it significantly reduces attack success rates across both trajectory and reasoning surfaces.

\section{Discussion and Conclusion}
We presented a systematic study of reasoning vulnerabilities in reasoning-enabled VLA models for autonomous driving, along with an evaluation pipeline that links textual perturbations to reasoning changes, trajectory degradation, and downstream safety impact. Our results show that realistic textual perturbations can manipulate semantic and structural reasoning properties, degrade trajectory behavior, and increase safety-critical failures. We further show that reasoning constitutes a distinct attack surface beyond trajectory prediction, and that lightweight input normalization substantially reduces attack success, providing an effective first line of protection against textual corruptions.

While our defense mitigates surface-level textual corruptions, it does not address adaptive attacks, semantic perturbations, or attacks in the visual input domain. Future work should develop stronger real-time defenses and study adversaries with partial or full model access, including visual-only and multimodal attacks that jointly target visual and textual inputs. Overall, our results call for systematic robustness evaluation before deploying reasoning-enabled embodied systems.

\bibliographystyle{plain}
\bibliography{main}
\appendix

\newpage

\section{Extended Open-Loop Results}

\begin{table}[t]
\centering
\caption{Open-loop ASR and downstream safety impact across reasoning and trajectory targets.}
\label{tab:open-loop_main-apdx}
\small
\setlength{\tabcolsep}{4pt}
\resizebox{0.9\textwidth}{!}{%
\begin{tabular}{@{}l l l l c c ccc@{}}
\toprule
\multirow{2}{*}{\textbf{Model}} 
& \multirow{2}{*}{\textbf{Surface}} 
& \multirow{2}{*}{\textbf{Objective}} 
& \multirow{2}{*}{\textbf{Target}}
& \multirow{2}{*}{\textbf{ASR} $\uparrow$}
& \multirow{2}{*}{\textbf{Queries} $\downarrow$}
& \multicolumn{3}{c}{\textbf{Safety Metrics (abs.\;($\Delta$))}} \\
\cmidrule(l){7-9}
&&&&&& Coll. (\%)  $\uparrow$ & N.-Enc. (\%) $\uparrow$ & TTC (ms) $\downarrow$ \\
\midrule

\multirow{9}{*}{Alp1}
& \multirow{7}{*}{Reasoning}
& \multirow{5}{*}{Semantic}
& Object      & 0.765 & 10.2 & 8.45 (-0.53) & 67.43 (0.14) & 950 (-36) \\
& & & Relation    & 0.889 &  6.8 & 7.29 (-0.79) & 66.55 (0.35) & 960 (-13) \\
& & & Implication & 0.850 &  7.6 & 7.84 (-0.60) & 68.76 (0.48) & 929 (-7) \\
& & & Planning    & 0.832 &  8.3 & 6.90 (-1.36) & 65.48 (0.62) & 999 (-27) \\
& & & Overall     & 0.836 &  9.4 & 7.74 (-0.85) & 68.10 (0.37) & 945 (-5) \\
\cmidrule(l){3-9}
& & \multirow{2}{*}{Structural}
& Slowdown & 0.248 & 31.9 & 6.69 (-2.40) & 72.31 (-0.87) & 989 (-125) \\
& & & DoS      & 0.047 & 37.4 &  8.89 (2.22) & 69.56 (-10.89) & 492 (101) \\
\cmidrule(l){2-9}
& Trajectory
& Deviation
& ADE & 0.336 & 18.6 & 15.56 (5.81) & 65.55 (4.57) & 1091 (-145) \\

\midrule

\multirow{9}{*}{Alp1.5}
& \multirow{7}{*}{Reasoning}
& \multirow{5}{*}{Semantic}
& Object      & 0.436 & 10.5 & 6.12 (1.41) & 69.91 (0.01) & 953 (-87) \\
& & & Relation    & 0.626 &  9.8 & 4.92 (0.00) & 72.31 (1.49) & 802 (-29) \\
& & & Implication & 0.520 &  9.8 & 7.11 (0.99) & 75.55 (0.79) & 716 (-44) \\
& & & Planning    & 0.429 & 10.5 & 4.78 (1.19) & 71.06 (-0.73) & 822 (-38) \\
& & & Overall     & 0.422 &  8.9 & 4.87 (0.00) & 71.55 (0.02) & 737 (-54) \\
\cmidrule(l){3-9}
& & \multirow{2}{*}{Structural}
& Slowdown & 0.102 & 12.1 & 10.12 (5.06) & 63.64 (0.00) & 962 (-72) \\
& & & DoS      & 0.088 & 15.8 & 11.58 (5.75) & 76.84 (1.11) & 653 (-17) \\
\cmidrule(l){2-9}
& Trajectory
& Deviation
& ADE & 0.115 & 21.5 & 10.68 (10.68) & 74.16 (9.70) & 1678 (-132) \\

\bottomrule
\end{tabular}
}
\end{table}

\begin{figure}
    \centering
    \includegraphics[width=0.6\linewidth]{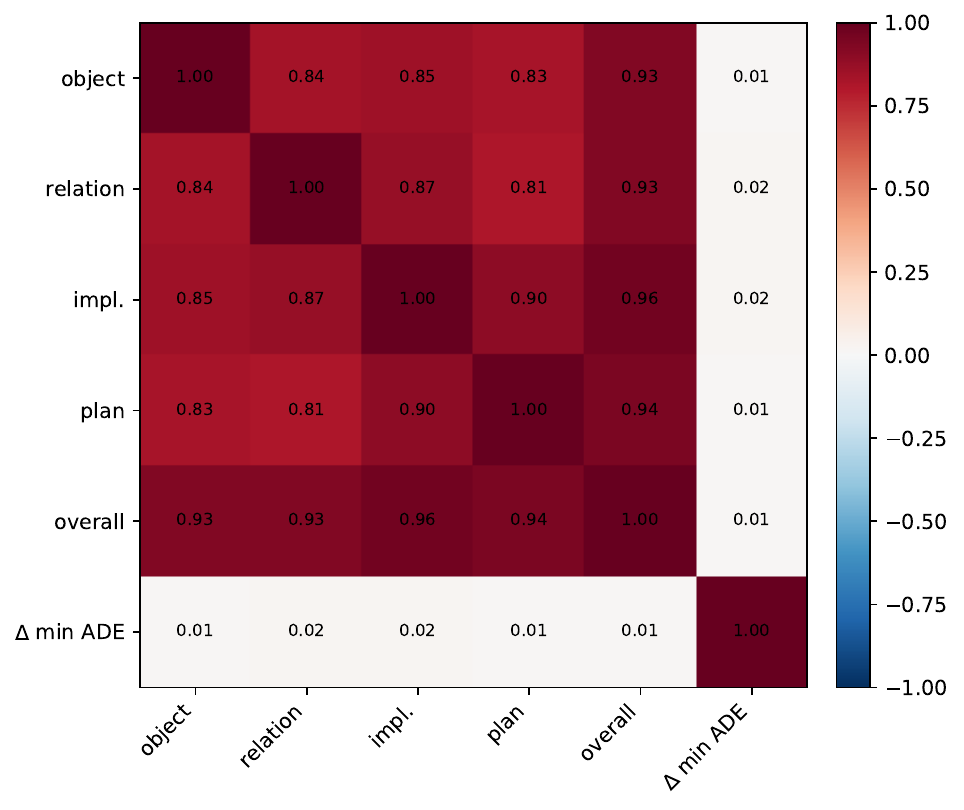}
    \caption{Alp1.5: Reasoning-trajectory correlation}
    \label{fig:correlation-a15}
\end{figure}

\subsection{Effect of Query Budget and Threshold}
\label{apdx:query_budget_threshold}

\begin{figure}[t]
\centering

\begin{subfigure}{0.45\linewidth}
    \centering
    \includegraphics[width=\linewidth]{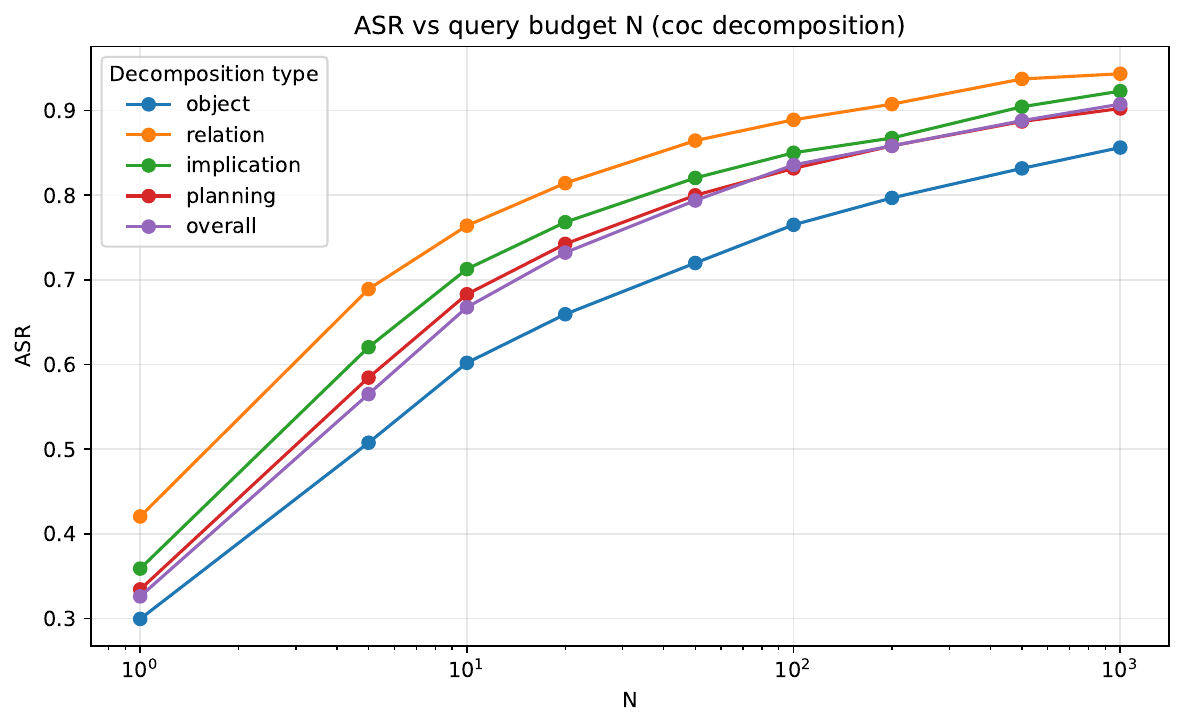}
    \caption{Target: CoC Semantic Decomposition}
\end{subfigure}
\hfill
\begin{subfigure}{0.45\linewidth}
    \centering
    \includegraphics[width=\linewidth]{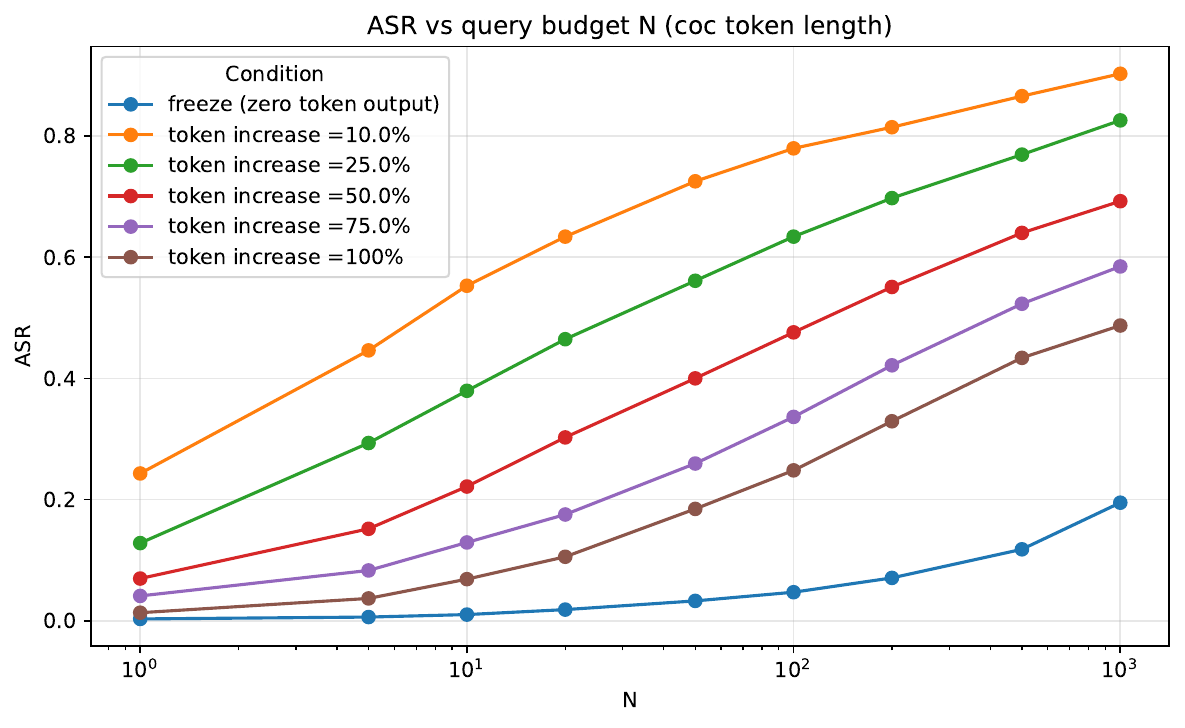}
    \caption{Target: CoC Token Length}
\end{subfigure}

\vspace{0.5em} 

\begin{subfigure}{0.45\linewidth}
    \centering
    \includegraphics[width=\linewidth]{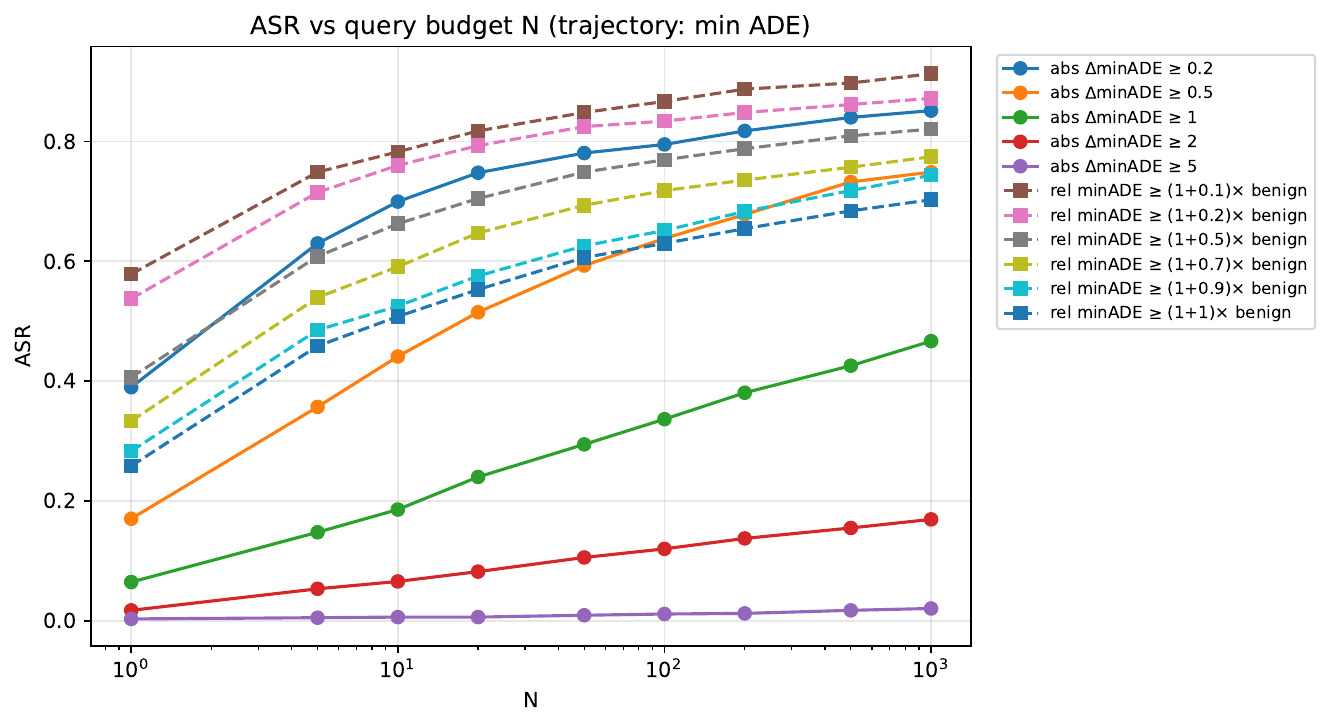}
    \caption{Target: Trajectory Prediction}
\end{subfigure}

\caption{Alpamayo1 open-loop ASR under varying query budgets and success thresholds across reasoning and trajectory objectives.}
\label{fig:a1_all}
\end{figure}

\begin{figure}[t]
\centering

\begin{subfigure}{0.45\linewidth}
    \centering
    \includegraphics[width=\linewidth]{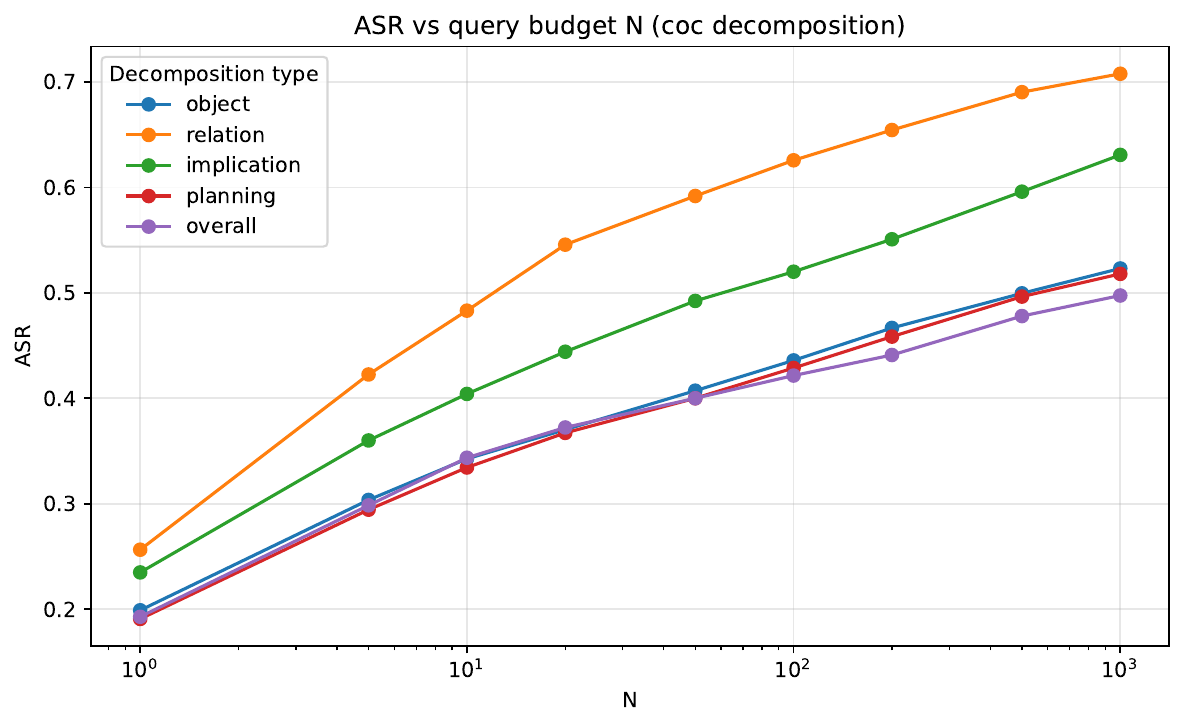}
    \caption{Target: CoC Semantic Decomposition}
\end{subfigure}
\hfill
\begin{subfigure}{0.45\linewidth}
    \centering
    \includegraphics[width=\linewidth]{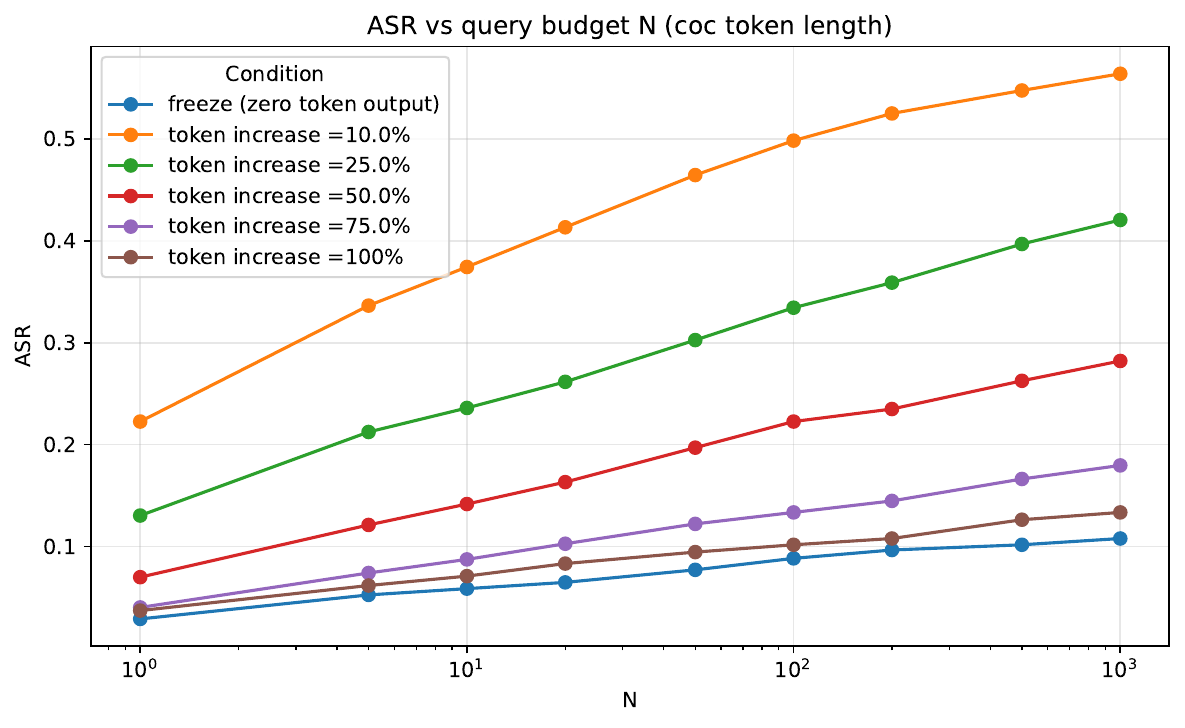}
    \caption{Target: CoC Token Length}
\end{subfigure}

\vspace{0.5em} 

\begin{subfigure}{0.45\linewidth}
    \centering
    \includegraphics[width=\linewidth]{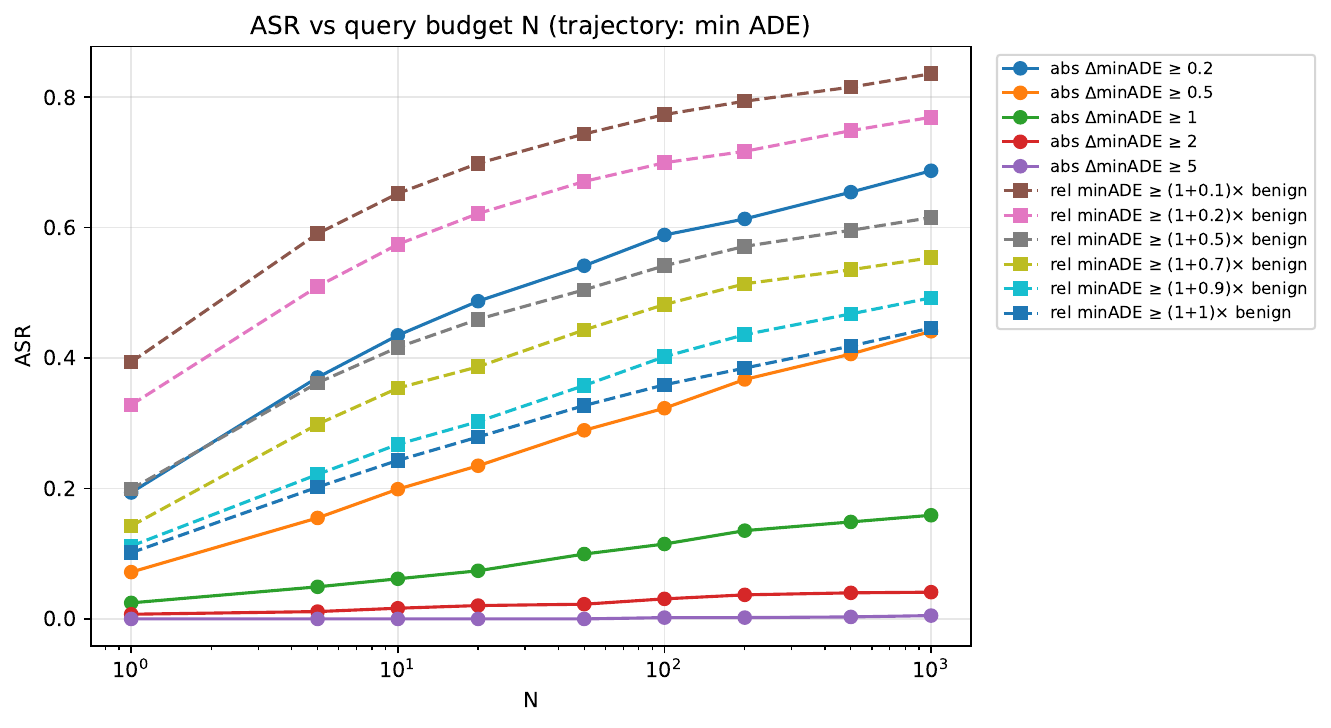}
    \caption{Target: Trajectory Prediction}
\end{subfigure}

\caption{Alpamayo1.5 open-loop ASR under varying query budgets and success thresholds across reasoning and trajectory objectives.}
\label{fig:a1.5_all}
\end{figure}

We further analyze how the query budget and success threshold affect attack success in the open-loop setting. 
Figures~\ref{fig:a1_all} and~\ref{fig:a1.5_all} report ASR as a function of the number of perturbation queries \(N\) for Alpamayo1 and Alpamayo1.5, respectively, across semantic reasoning, structural reasoning, and trajectory objectives. 
Unless otherwise stated, our main open-loop experiments use \(N=100\) as the default query budget, \(\delta_{\text{traj}}=1\) for trajectory manipulation based on min-ADE, and a \(100\%\) token increase threshold for slowdown, i.e., the perturbed reasoning must be at least twice as long as the benign reasoning. 

We choose \(N=100\) because it provides a practical trade-off: it is not an excessively large query budget, yet it achieves substantial ASR and lies near the saturation region of most curves. 
For trajectory manipulation, we use a 1m min-ADE threshold because it captures physically meaningful trajectory deviation while still yielding measurable attack success. 
For slowdown, we use the stricter \(100\%\) token-increase threshold to test whether perturbations can substantially push the model into longer reasoning generation.

Across both models, ASR increases monotonically with the query budget, showing that the perturbation space contains many successful variants that become easier to discover as more queries are allowed. 
This trend is strongest for semantic reasoning objectives: object, relation, implication, planning, and overall reasoning deviation all improve steadily with larger \(N\). 
Alpamayo1 is consistently more vulnerable than Alpamayo1.5, reaching higher ASR under the same query budget, which is consistent with the stronger robustness of Alpamayo1.5 observed in the main results.

Threshold choice also has a substantial effect on the measured ASR. 
For structural reasoning attacks, lower token-increase thresholds yield higher success rates, while stricter slowdown criteria reduce ASR but remain non-negligible at larger query budgets. 
The freeze/zero-token condition remains the hardest structural objective, indicating that fully suppressing reasoning is substantially more difficult than increasing reasoning length. 
For trajectory attacks, smaller absolute or relative min-ADE thresholds produce higher ASR, while stricter thresholds require larger query budgets and remain harder to satisfy. 
This confirms that textual perturbations often induce measurable trajectory drift, but large trajectory deviations require either more favorable perturbations or a larger query budget.

Overall, these results show that attack success is not an artifact of a single threshold choice. 
Instead, both reasoning and trajectory vulnerabilities persist across a range of thresholds, with ASR increasing smoothly as the adversary is allowed more queries. 
At the same time, the results highlight the importance of reporting query budgets and thresholds explicitly: overly permissive thresholds can overstate practical risk, while overly strict thresholds may hide smaller deviations that can still compound in closed-loop driving.

\subsection{Malicious vs. Neutral Inputs} \label{apdx:mal_vs_neutral}

\begin{table}[t]
\centering
\caption{Effect of clean malicious prompts without textual augmentation. Results compare neutral inputs against malicious prompts with different semantic intents and prompt types at \(N=0\).}
\label{tab:mal_benign_semantic}
\small
\setlength{\tabcolsep}{5pt}
\renewcommand{\arraystretch}{1.15}

\begin{tabular}{lllcccccc}
\toprule
\multirow{4}{*}{\makecell{\textbf{Prompt}\\\textbf{Semantic}}}
& \multirow{4}{*}{\textbf{Intent}}
& \multirow{4}{*}{\textbf{Type}}
& \textbf{Trajectory}
& \multicolumn{5}{c}{\textbf{Reasoning}} \\
\cmidrule(lr){4-4} \cmidrule(lr){5-9}
& & & \multirow{2}{*}{\textbf{min ADE}}
& \multicolumn{5}{c}{\textbf{Semantic}} \\
\cmidrule(lr){5-9}
& & &
& \textbf{Object} & \textbf{Relation} & \textbf{Impl.}
& \textbf{Planning} & \textbf{Overall} \\
\midrule

Neutral
& -- & -- & 0.93 & -- & -- & -- & -- & -- \\

\midrule

\multirow{10}{*}{Malicious}
& \multirow{2}{*}{Object}
& Obvious   & 0.84 & 0.11 & 0.16 & 0.15 & 0.10 & 0.14 \\
& & Plausible & 0.86 & 0.09 & 0.15 & 0.15 & 0.08 & 0.12 \\
\cmidrule(lr){2-9}

& \multirow{2}{*}{Relation}
& Obvious   & 0.87 & 0.09 & 0.15 & 0.14 & 0.08 & 0.12 \\
& & Plausible & 0.88 & 0.09 & 0.17 & 0.16 & 0.09 & 0.14 \\
\cmidrule(lr){2-9}

& \multirow{2}{*}{Impl.}
& Obvious   & 0.86 & 0.10 & 0.16 & 0.15 & 0.09 & 0.13 \\
& & Plausible & 0.86 & 0.11 & 0.18 & 0.16 & 0.10 & 0.14 \\
\cmidrule(lr){2-9}

& \multirow{2}{*}{Planning}
& Obvious   & 0.86 & 0.10 & 0.16 & 0.15 & 0.08 & 0.12 \\
& & Plausible & 0.85 & 0.13 & 0.17 & 0.15 & 0.10 & 0.14 \\
\cmidrule(lr){2-9}

& \multirow{2}{*}{Overall}
& Obvious   & 0.87 & 0.10 & 0.16 & 0.16 & 0.10 & 0.13 \\
& & Plausible & 0.86 & 0.14 & 0.19 & 0.15 & 0.11 & 0.15 \\

\bottomrule
\end{tabular}
\end{table}

\begin{table}[t]
\centering
\caption{Attack success rates for neutral versus malicious inputs under textual augmentation. ASR is reported across trajectory, semantic reasoning, and structural reasoning objectives.}
\label{tab:mal_benign_semantic_asr}
\small
\setlength{\tabcolsep}{4pt}
\begin{tabular}{@{}l l l l c c ccc@{}}
\toprule
\multirow{2}{*}{\makecell{\textbf{Prompt}\\\textbf{Semantic}}}
& \multirow{2}{*}{\textbf{Surface}} 
& \multirow{2}{*}{\textbf{Objective}} 
& \multirow{2}{*}{\textbf{Target}}
& \multirow{2}{*}{\textbf{ASR} $\uparrow$}
& \multirow{2}{*}{\textbf{Queries} $\downarrow$}
& \multicolumn{3}{c}{\textbf{Safety Metrics (abs.\;($\Delta$))}} \\
\cmidrule(l){7-9}
&&&&&& Coll. (\%)  $\uparrow$ & N.-Enc. (\%) $\uparrow$ & TTC (ms) $\downarrow$ \\
\midrule

\multirow{9}{*}{Neutral}
& \multirow{7}{*}{Reasoning}
& \multirow{5}{*}{Semantic}
& Object      & 0.765 & 10.2 & 8.45 (-0.53) & 67.43 (0.14) & 950 (-36) \\
& & & Relation    & 0.889 &  6.8 & 7.29 (-0.79) & 66.55 (0.35) & 960 (-13) \\
& & & Implication & 0.850 &  7.6 & 7.84 (-0.60) & 68.76 (0.48) & 929 (-7) \\
& & & Planning    & 0.832 &  8.3 & 6.90 (-1.36) & 65.48 (0.62) & 999 (-27) \\
& & & Overall     & 0.836 &  9.4 & 7.74 (-0.85) & 68.10 (0.37) & 945 (-5) \\
\cmidrule(l){3-9}
& & \multirow{2}{*}{Structural}
& Slowdown & 0.248 & 31.9 & 6.69 (-2.40) & 72.31 (-0.87) & 989 (-125) \\
& & & DoS      & 0.047 & 37.4 &  8.89 (2.22) & 69.56 (-10.89) & 492 (101) \\
\cmidrule(l){2-9}
& Trajectory
& Deviation
& ADE & 0.336 & 18.6 & 15.56 (5.81) & 65.55 (4.57) & 1091 (-145) \\

\midrule

\multirow{9}{*}{Malicious}
& \multirow{7}{*}{Reasoning}
& \multirow{5}{*}{Semantic}
& Object     & 0.772 & 9.26 & 9.43 (0.13) & 69.06 (0.93) & 1009 (12) \\
& & & Relation    & 0.896 & 6.85 & 7.89 (-0.12) & 67.51 (1.03) & 1038 (5) \\
& & & Implication & 0.866 & 7.96 & 6.64 (-1.65) & 68.13 (0.12) & 1020 (30) \\
& & & Planning    & 0.829 & 8.083 & 6.93 (-1.36) & 67.70 (0.62) & 1035 (52) \\
& & & Overall      & 0.818 & 7.863 & 8.40 (-0.38) & 68.04 (0.50) & 1042 (27) \\
\cmidrule(l){3-9}
& & \multirow{2}{*}{Structural}
& Slowdown & 0.297 & 32.97 & 9.64 (-1.37) & 73.78 (-2.72) & 817 (-161) \\
& & & DoS     & 0.058 & 42.70 & 7.62 (-7.91) & 69.02 (-9.44) & 1376 (-2) \\
\cmidrule(l){2-9}
& Trajectory
& Deviation
& ADE & 0.35 & 19.29 & 16.97 (7.00) & 65.65 (3.82) & 1107 (-42) \\

\bottomrule
\end{tabular}
\end{table}

\begin{table*}[t]
\centering
\caption{ASR by malicious prompt semantic intent and prompt type across trajectory, semantic reasoning, and structural objectives.}
\label{tab:mal_semantic_comp}
\small
\setlength{\tabcolsep}{4pt}

\begin{tabular}{lccccccccc}
\toprule
\multirow{4}{*}{\textbf{Intent}} 
& \multirow{4}{*}{\textbf{Type}}
& \multirow{3}{*}{\textbf{Trajectory}}
& \multicolumn{7}{c}{\textbf{Reasoning}} \\

\cmidrule(lr){4-10}

& & & \multicolumn{5}{c}{Semantic}
& \multicolumn{2}{c}{Structural} \\

\cmidrule(lr){3-3} \cmidrule(lr){4-8} \cmidrule(lr){9-10}

& & ADE & Object & Relation & Impl. & Planning & Overall
& Slowdown & DoS \\
\midrule
\multirow{2}{*}{Object} & Obvious & 0.376 & 0.770 & 0.892 & 0.874 & 0.855 & 0.838 & 0.292 & 0.054 \\
 & Plausible & 0.384 & 0.782 & 0.889 & 0.862 & 0.836 & 0.848 & 0.268 & 0.058 \\
\cmidrule(l){1-10}
\multirow{2}{*}{Relation} & Obvious & 0.353 & 0.762 & 0.867 & 0.853 & 0.837 & 0.817 & 0.285 & 0.041 \\
 & Plausible & 0.352 & 0.775 & 0.889 & 0.857 & 0.847 & 0.839 & 0.243 & 0.064 \\
\cmidrule(l){1-10}
\multirow{2}{*}{Impl.} & Obvious & 0.363 & 0.776 & 0.902 & 0.870 & 0.844 & 0.837 & 0.313 & 0.069 \\
 & Plausible & 0.362 & 0.760 & 0.894 & 0.838 & 0.835 & 0.817 & 0.271 & 0.054 \\
\cmidrule(l){1-10}
\multirow{2}{*}{Planning} & Obvious & 0.344 & 0.790 & 0.897 & 0.871 & 0.828 & 0.835 & 0.262 & 0.053 \\
 & Plausible & 0.367 & 0.824 & 0.906 & 0.824 & 0.829 & 0.831 & 0.259 & 0.056 \\
\cmidrule(l){1-10}
\multirow{2}{*}{Universal} & Obvious & 0.350 & 0.772 & 0.896 & 0.866 & 0.829 & 0.818 & 0.297 & 0.058 \\
 & Plausible & 0.356 & 0.821 & 0.892 & 0.812 & 0.840 & 0.825 & 0.258 & 0.068 \\

\bottomrule
\end{tabular}
\end{table*}

We further compare semantically neutral inputs with semantically malicious inputs to separate the effect of malicious intent from the effect of textual corruption. 
Table~\ref{tab:mal_benign_semantic} reports results for clean malicious prompts without augmentation (\(N=0\)). 
Across different semantic intents and prompt types, clean malicious prompts produce only small deviations in reasoning and do not substantially change the trajectory error compared to the neutral input, even slightly decreasing it. 
This suggests that the model is relatively robust to explicit malicious semantics when the input is well-formed: the model often ignores the malicious instruction rather than directly following it.

However, once realistic textual corruptions are applied, the difference between neutral and malicious inputs becomes small. 
As shown in Table~\ref{tab:mal_benign_semantic_asr}, attacks starting from malicious inputs achieve ASR values comparable to those starting from neutral inputs across trajectory, semantic reasoning, and structural reasoning objectives. 
For example, trajectory ASR changes only slightly, while semantic reasoning ASR remains in a similar range across object, relation, implication, planning, and overall targets. 
Structural attacks also show only modest differences, with slowdown and DoS remaining harder than semantic reasoning manipulation.

Table~\ref{tab:mal_semantic_comp} further breaks down malicious inputs by semantic intent and prompt type. 
The results show no single malicious intent or prompt type consistently dominates across all objectives. 
Obvious and plausible malicious variants lead to similar ASR values, and the differences across object-, relation-, implication-, planning-, and universal-intent prompts are relatively small. 
Therefore, the attack effectiveness is not primarily driven by a specific malicious instruction template.

\subsection{Effect of Navigation Commands}
\label{apdx:nav}

Alpamayo1.5 additionally accepts navigation commands as part of its textual input, providing another potential input channel for perturbation. 
We therefore evaluate whether perturbing this auxiliary navigation input leads to vulnerabilities similar to those observed for the main textual input. 
Table~\ref{tab:nav_results} reports the corresponding ASR and safety impact across trajectory, semantic reasoning, and structural reasoning objectives.

The results show that navigation commands are also vulnerable to realistic textual perturbations. 
Nav-based attacks achieve non-trivial ASR across both reasoning and trajectory objectives, indicating that the model does not rely only on the primary instruction channel when forming its reasoning and trajectory outputs. 
In particular, perturbations to navigation commands can alter semantic reasoning components and induce trajectory deviations, showing that auxiliary textual inputs can propagate through both output surfaces.

The resulting safety impact is also comparable to the trends observed for the main textual input. 
Successful navigation-command attacks can increase collision and near-encounter rates and reduce min-TTC, suggesting that perturbing this auxiliary channel can still lead to safety-relevant degradation. 

Overall, Table~\ref{tab:nav_results} shows that vulnerabilities are not limited to a single textual field. 
For Alpamayo1.5, both the main textual input and navigation commands can be exploited through realistic surface-form corruptions.

\begin{table}[t]
\centering
\caption{Effect of perturbing Alpamayo1.5 navigation commands. ASR is reported across trajectory, semantic reasoning, and structural reasoning objectives; safety metrics are reported on successful attacks.}
\label{tab:nav_results}
\small
\setlength{\tabcolsep}{4pt}
\begin{tabular}{@{}l l l l c c ccc@{}}
\toprule
\multirow{2}{*}{\makecell{\textbf{Prompt} \\ \textbf{Type}}}
& \multirow{2}{*}{\textbf{Surface}} 
& \multirow{2}{*}{\textbf{Objective}} 
& \multirow{2}{*}{\textbf{Target}}
& \multirow{2}{*}{\textbf{ASR} $\uparrow$}
& \multirow{2}{*}{\textbf{Queries} $\downarrow$}
& \multicolumn{3}{c}{\textbf{Safety Metrics (abs.\;($\Delta$))}} \\
\cmidrule(l){7-9}
&&&&&& Coll. (\%)  $\uparrow$ & N.-Enc. (\%) $\uparrow$ & TTC (ms) $\downarrow$ \\
\midrule

\multirow{9}{*}{\makecell{Sys \\ \& User}}
& \multirow{7}{*}{Reasoning}
& \multirow{5}{*}{Semantic}
& Object      & 0.436 & 10.5 & 6.12 (1.41) & 69.91 (0.01) & 953 (-87) \\
& & & Relation    & 0.626 &  9.8 & 4.92 (0.00) & 72.31 (1.49) & 802 (-29) \\
& & & Implication & 0.520 &  9.8 & 7.11 (0.99) & 75.55 (0.79) & 716 (-44) \\
& & & Planning    & 0.429 & 10.5 & 4.78 (1.19) & 71.06 (-0.73) & 822 (-38) \\
& & & Overall     & 0.422 &  8.9 & 4.87 (0.00) & 71.55 (0.02) & 737 (-54) \\
\cmidrule(l){3-9}
& & \multirow{2}{*}{Structural}
& Slowdown & 0.102 & 12.1 & 10.12 (5.06) & 63.64 (0.00) & 962 (-72) \\
& & & DoS      & 0.088 & 15.8 & 11.58 (5.75) & 76.84 (1.11) & 653 (-17) \\
\cmidrule(l){2-9}
& Trajectory
& Deviation
& ADE & 0.115 & 21.5 & 10.68 (10.68) & 74.16 (9.70) & 1678 (-132) \\

\midrule

\multirow{9}{*}{Nav}
& \multirow{7}{*}{Reasoning}
& \multirow{5}{*}{Semantic}
& Object      & 0.400 & 11.8 & 6.67 (1.54) & 74.89 (2.05) & 786 (35) \\
& & & Relation    & 0.427 & 11.7 & 7.46 (1.45) & 75.73 (0.97) & 749 (54) \\
& & & Implication & 0.375 & 12.90 & 6.01 (0.54) & 75.68 (0.26) & 735 (81) \\
& & & Planning    & 0.381 & 11.32 & 5.12 (-0.27) & 69.27 (0.26) & 885 (75) \\
& & & Overall    & 0.363 & 10.44 & 6.78 (1.13) & 74.00 (0.83) & 836 (68) \\
\cmidrule(l){3-9}
& & \multirow{2}{*}{Structural}
& Slowdown  & 0.105 & 19.85 & 18.67 (3.95) & 86.29 (-3.90) & 412 (-40) \\
& & & DoS      & 0.023 & 36.1 & 0.00 (0.00) & 91.00 (14.00) & 1006 (-371) \\
\cmidrule(l){2-9}
& Trajectory
& Deviation
& ADE & 0.124 & 12.65 & 8.27 (8.27) & 78.53 (4.97) & 1098 (-134) \\

\bottomrule
\end{tabular}
\end{table}

\subsection{Effect of Augmentation Strength}  \label{apdx:aug}
We analyze how perturbation intensity affects attack success by varying the augmentation strength \(\sigma\). 
Following prior work, we use \(\sigma=0.4\) as the default setting in our main experiments, as it provides a practical middle ground: it is strong enough to reveal model vulnerabilities, but not so aggressive that it destroys the readability or intended meaning of the input text. 
As shown in Table~\ref{tab:aug}, ASR generally increases with larger \(\sigma\) across trajectory, semantic reasoning, and structural reasoning objectives. 
However, even mild perturbations remain effective, with \(\sigma=0.1\) already producing substantial semantic reasoning ASR. 
This indicates that the observed vulnerabilities are not only an artifact of extreme textual corruption, while stronger perturbations naturally expose a larger attack surface.

\begin{table*}[t]
\centering
\caption{Effect of augmentation strength on open-loop ASR. Larger \(\sigma\) corresponds to stronger textual perturbations.}
\label{tab:aug}
\small
\setlength{\tabcolsep}{5pt}
\renewcommand{\arraystretch}{1.2}

\begin{tabular}{lcccccccc}
\toprule
\multirow{4}{*}{\textbf{$\sigma$}} 
& \multirow{3}{*}{\textbf{Trajectory}}
& \multicolumn{7}{c}{\textbf{Reasoning}} \\

\cmidrule(lr){3-9}

& & \multicolumn{5}{c}{Semantic}
& \multicolumn{2}{c}{Structural} \\

\cmidrule(lr){2-2} \cmidrule(lr){3-7} \cmidrule(lr){8-9}

& ADE & Object & Relation & Impl. & Planning & Overall
& Slowdown & DoS \\
\midrule

\textbf{0.1} & 0.248 & 0.664 & 0.814 & 0.656 & 0.690 & 0.687 & 0.137 & 0.011 \\
\textbf{0.25} & 0.309 & 0.764 & 0.889 & 0.781 & 0.770 & 0.772 & 0.202 & 0.031 \\
\textbf{0.4} & 0.336 & 0.765 & 0.889 & 0.850 & 0.832 & 0.836 & 0.248 & 0.047 \\
\textbf{0.7} & 0.392 & 0.851 & 0.953 & 0.882 & 0.856 & 0.867 & 0.317 & 0.069 \\

\bottomrule
\end{tabular}
\end{table*}

\subsection{Effect of Input Token Length}  \label{apdx:input_token_length}
We further study whether longer textual inputs increase vulnerability under perturbation. 
In our main experiments, we use 29 tokens because this matches the default prompt length used by the evaluated models. 
Table~\ref{tab:input_token_length} shows that ASR increases substantially as input length grows from 12 to 51 tokens, especially for trajectory deviation and semantic reasoning manipulation. 
This suggests that longer inputs provide a larger perturbable surface: even when each individual corruption is small, more tokens create more opportunities for perturbations to affect the model output. 
Structural attacks show a weaker but still generally increasing trend, with DoS becoming more frequent for longer inputs.

\begin{table*}[t]
\centering
\caption{Effect of input token length on open-loop ASR. Longer textual inputs provide a larger perturbable surface and generally increase vulnerability.}
\label{tab:input_token_length}
\small
\setlength{\tabcolsep}{5pt}
\renewcommand{\arraystretch}{1.2}

\begin{tabular}{ccccccccc}
\toprule
\multirow{4}{*}{\textbf{Token Length}} 
& \multirow{3}{*}{\textbf{Trajectory}}
& \multicolumn{7}{c}{\textbf{Reasoning}} \\

\cmidrule(lr){3-9}

& & \multicolumn{5}{c}{Semantic}
& \multicolumn{2}{c}{Structural} \\

\cmidrule(lr){2-2} \cmidrule(lr){3-7} \cmidrule(lr){8-9}

& ADE & Object & Relation & Impl. & Planning & Overall
& Slowdown & DoS \\
\midrule

\textbf{12} & 0.154 & 0.521 & 0.679 & 0.507 & 0.484 & 0.510 & 0.135 & 0.017 \\
\textbf{29} & 0.336 & 0.765 & 0.889 & 0.850 & 0.832 & 0.836 & 0.248 & 0.047 \\
\textbf{51} & 0.530 & 0.891 & 0.955 & 0.906 & 0.865 & 0.905 & 0.227 & 0.073 \\

\bottomrule
\end{tabular}
\end{table*}
\section{Extended Closed-Loop Analysis}

\subsection{Safety Impact}
\label{apdx:closed_safety}

\begin{figure}
    \centering
    \includegraphics[width=1.0\linewidth]{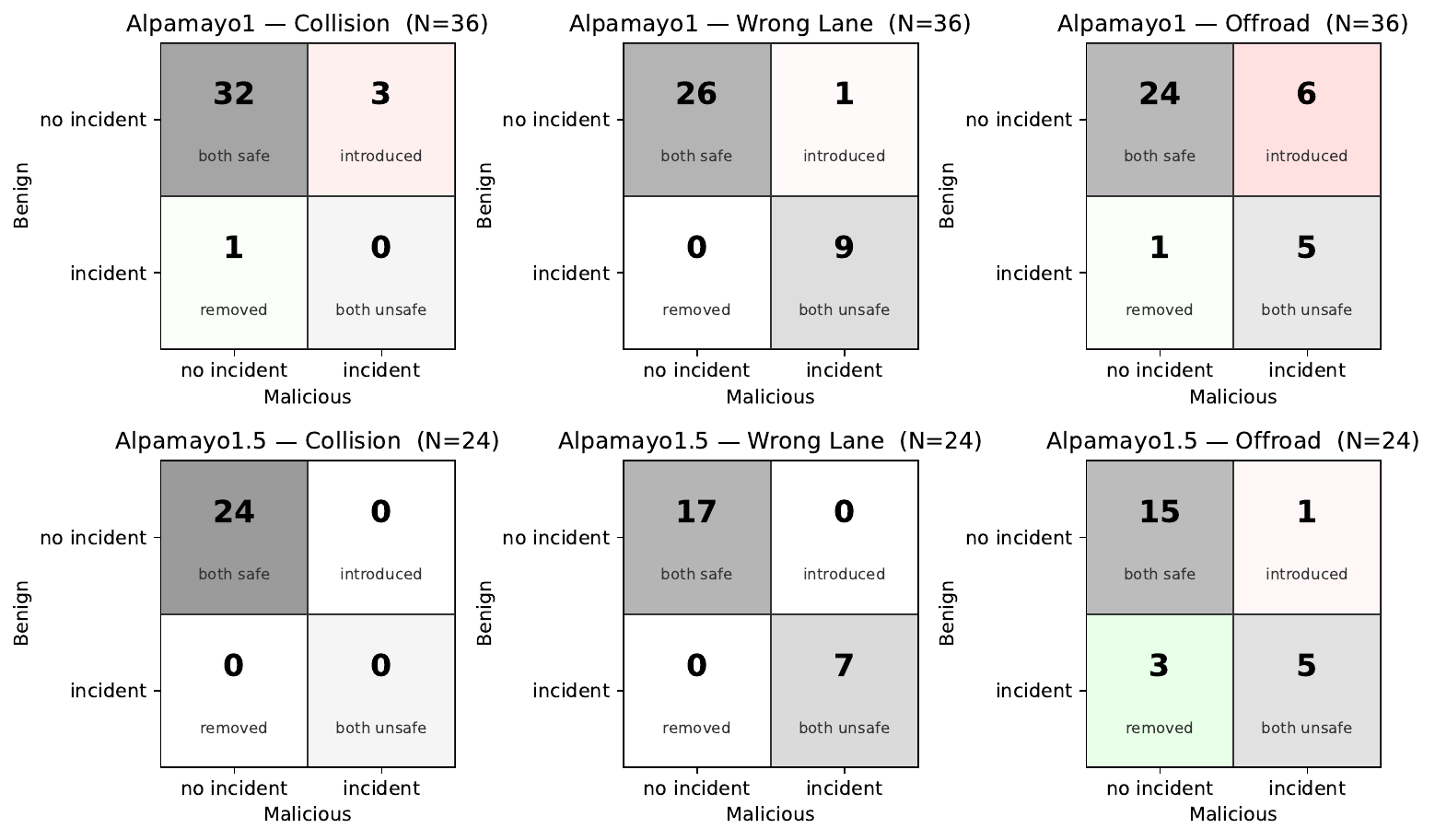}
    \caption{Closed-loop safety impact of successful trajectory-targeted attacks. Bars compare benign and perturbed rollouts across collision, off-road, and wrong-lane incidents for Alpamayo1 and Alpamayo1.5.}
    \label{fig:closed_traj_safety}
\end{figure}

\begin{figure}
    \centering
    \includegraphics[width=1.0\linewidth]{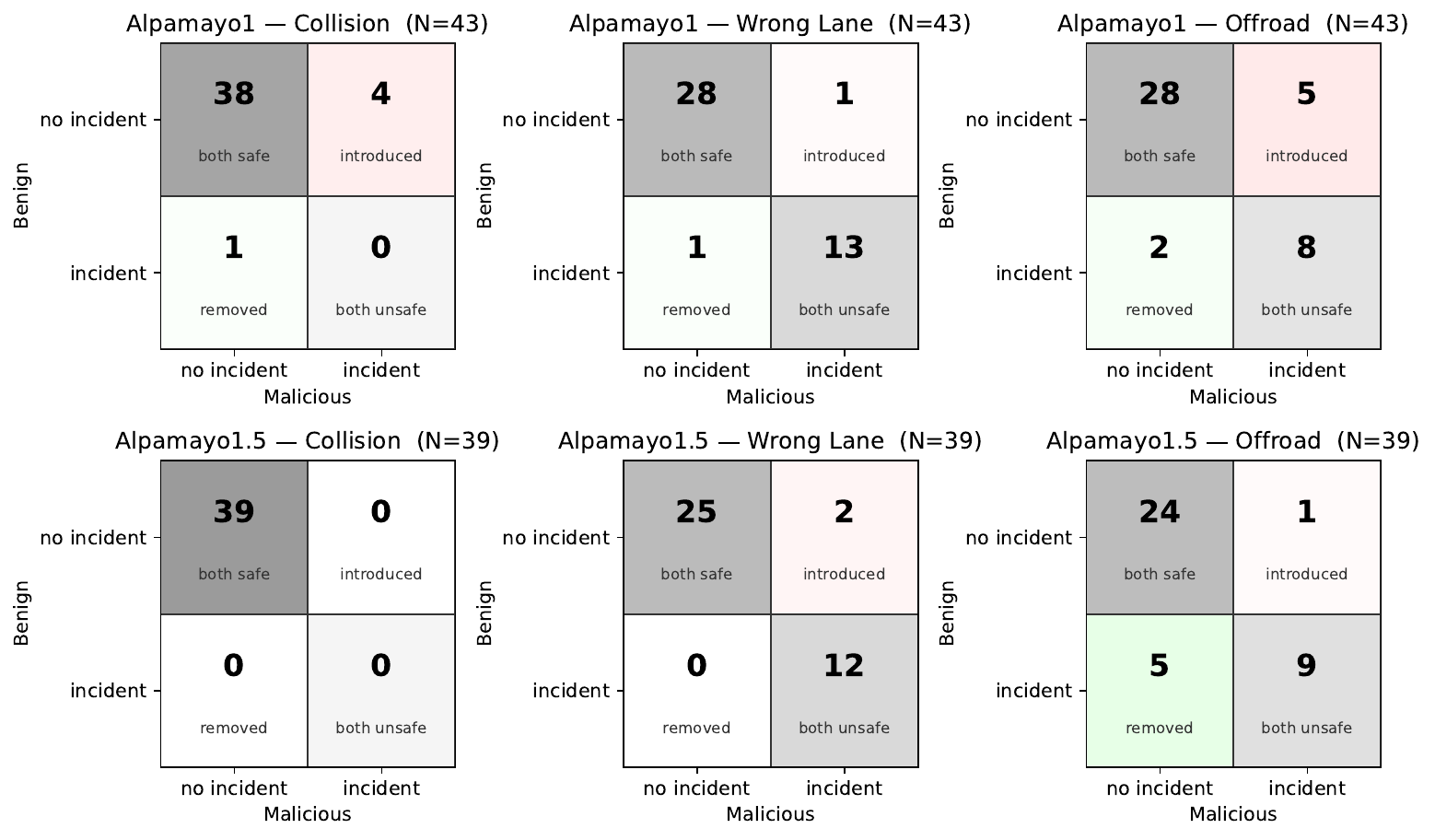}
   \caption{Closed-loop safety impact of successful reasoning-targeted attacks. Bars compare benign and perturbed rollouts across collision, off-road, and wrong-lane incidents for Alpamayo1 and Alpamayo1.5.}
\label{fig:closed_reas_safety}
\end{figure}

We further analyze the downstream safety impact of successful closed-loop attacks. 
Figures~\ref{fig:closed_traj_safety} and~\ref{fig:closed_reas_safety} compare the incident rates of benign and perturbed rollouts for trajectory- and reasoning-targeted attacks, respectively. 
Overall, corrupted textual inputs make model behavior less stable: perturbations often push the rollout away from its nominal behavior, and these deviations can introduce new safety violations such as collisions, off-road events, or wrong-lane driving.

For Alpamayo1, this instability leads to clear safety degradation. 
In particular, successful perturbations increase the number of collision incidents by several cases, with approximately three to four additional collisions depending on the attack surface. 

The effect is not uniform across all incident types, however. 
When the benign rollout is already unsafe, perturbations may occasionally move the vehicle away from the original failure mode and reduce a particular incident. 
Such cases should not be interpreted as attack improvement; rather, they reflect the high sensitivity of closed-loop driving dynamics, where deviations from an already unstable trajectory can sometimes mask one failure while introducing or shifting another.

Alpamayo1.5 is relatively more robust, consistent with the lower closed-loop ASR reported in the main results. 
Nevertheless, it remains vulnerable: successful perturbations still induce non-trivial safety changes and can destabilize rollout behavior. 
A notable observation is that Alpamayo1.5 already exhibits a high off-road rate under benign execution. 
Because the nominal behavior is itself unstable for this metric, some perturbations reduce off-road incidents relative to the benign rollout. 
This does not imply robustness or safety improvement under attack; instead, it shows that off-road behavior is dominated by a weak benign baseline in these scenarios, making the measured attack effect less clean than for collision or wrong-lane incidents.

Taken together, Figures~\ref{fig:closed_traj_safety} and~\ref{fig:closed_reas_safety} show that closed-loop safety impact is incident- and model-dependent. 
Corrupted inputs generally destabilize the model and often introduce new failures, especially for Alpamayo1, while in cases where the benign model already fails, perturbations can occasionally alter the failure mode or reduce a specific incident count. 
This reinforces the need to evaluate attacks in closed loop: the safety consequence of a reasoning or trajectory deviation cannot be fully captured by open-loop error alone.

\subsection{Pre-Incident Behavior Analysis}
\label{apdx:closed_incident_window_analysis}

To better understand how closed-loop failures emerge, we analyze model behavior in a short temporal window before each safety incident. 
Table~\ref{tab:closed_incident_window} reports the average trajectory deviation and reasoning deviation measured immediately before collision, wrong-lane, and off-road events. 
This analysis complements the scenario-level safety results in Appendix~\ref{apdx:closed_safety} by examining which output surface changes most strongly near failure onset.

The results show that closed-loop incidents are not associated with a single uniform failure channel. 
For some incident types, trajectory deviation is more prominent before failure, suggesting that corrupted inputs directly destabilize the realized ego motion. 
For other incidents, reasoning deviation is more pronounced, indicating that the model's intermediate reasoning can shift substantially near unsafe behavior even when trajectory deviation alone does not fully explain the event. 
This supports our broader observation that reasoning and trajectory interact in an incident-dependent manner under closed-loop dynamics.

Importantly, this analysis should not be interpreted as establishing causal attribution. 
Because closed-loop rollouts are dynamic and compounding, a perturbation may first affect reasoning, trajectory, or both, and later deviations can be amplified by simulator feedback. 
Therefore, Table~\ref{tab:closed_incident_window} should be viewed as a diagnostic analysis of pre-incident behavior rather than a definitive explanation of why each incident occurs. 
Nevertheless, the results show that successful attacks often produce measurable deviations before safety violations, reinforcing that corrupted textual inputs can destabilize both intermediate reasoning and downstream control.

\begin{table*}[t]
\centering
\caption{Trajectory and reasoning deviations measured before closed-loop incident onset.}
\label{tab:closed_incident_window}
\small
\setlength{\tabcolsep}{5pt}
\renewcommand{\arraystretch}{1.2}

\begin{tabular}{llccccc}
\toprule
\multirow{2}{*}{\textbf{Model}} 
& \multirow{2}{*}{\textbf{Incident}} 
& \multirow{2}{*}{\textbf{N. Clips}} 
& \multicolumn{2}{c}{\textbf{Deviation}} 
& \multicolumn{2}{c}{\textbf{ASR}} \\
\cmidrule(lr){4-5}
\cmidrule(lr){6-7}

& & 
& \textbf{Trajectory} 
& \textbf{Reasoning} 
& \textbf{Trajectory} 
& \textbf{Reasoning} \\

\midrule

\multirow{3}{*}{Alpamayo1}
& Collision  & 4 & 0.67 & 0.39 & 50 & 50 \\
& Wrong Lane & 1 & 1.12 & 0.04 & 100 & 0 \\
& Off-road   & 6 & 0.03 & 0.51 & 0 & 50 \\

\midrule

\multirow{3}{*}{Alpamayo1.5}
& Collision  & 0 & - & - & - & - \\
& Wrong Lane & 2 & -2.12 & 0.53 & 0 & 50 \\
& Off-road   & 2 & -0.80 & -0.35 & 0 & 0 \\

\bottomrule
\end{tabular}
\end{table*}

\subsection{Timing}
\label{apdx:closed_timing}

We analyze when safety incidents first occur during closed-loop rollouts. 
Table~\ref{tab:closed_timing} reports the average first time step of each incident type under benign and perturbed executions, where smaller values indicate earlier failure onset.

The results show that corrupted inputs can shift failures earlier in the rollout, especially for wrong-lane and off-road incidents. 
This suggests that perturbations do not only change whether a failure occurs, but can also accelerate the emergence of unsafe behavior. 
The effect is stronger for Alpamayo1, consistent with its higher closed-loop vulnerability, while Alpamayo1.5 remains relatively more stable but not immune.

These timing results should be interpreted together with the benign incident rates. 
When the benign rollout is already unstable, perturbations may sometimes delay or alter a specific incident type rather than uniformly making all failures earlier. 
Overall, Table~\ref{tab:closed_timing} shows that corrupted inputs affect both the occurrence and temporal onset of closed-loop failures.

\begin{table}[ht]
\centering
\caption{Average first occurrence time step of closed-loop incidents under benign and perturbed rollouts.}
\label{tab:closed_timing}
\small
\setlength{\tabcolsep}{6pt}
\renewcommand{\arraystretch}{1.2}
\begin{tabular}{l l r r r}
\toprule
\textbf{Model} & \textbf{Incident} 
& \textbf{Avg Second in Benign} 
& \textbf{Avg Second in Malicious} 
& \textbf{$\Delta$ (s)} \\
\midrule

\multirow{3}{*}{Alpamayo 1}
& Collision    & - & - & -  \\
& Wrong Lane   & 6.31  & 5.72  & -0.59  \\
& Off-road     & 7.43  & 6.12  & -1.31  \\

\midrule

\multirow{3}{*}{Alpamayo 1.5}
& Collision    & - & - & -  \\
& Wrong Lane   & 6.11 & 5.57 & -0.54 \\
& Off-road     & 5.85 & 5.42 & -0.43 \\
\bottomrule
\end{tabular}
\end{table}

\subsection{Different Window Size and Threshold}
\label{apdx:window_threshold}

We evaluate the sensitivity of closed-loop ASR to the temporal consistency requirement \(k\) and the deviation threshold \(\epsilon\). 
For trajectory deviation, Table~\ref{tab:asr_sensitivity_combined} shows that ASR is stable across moderate values of \(k\): for both models, increasing \(k\) from 1 to 5 changes ASR only slightly. 
This suggests that successful trajectory attacks are not dominated by isolated one-step spikes, but often persist over multiple consecutive steps. 
We therefore use \(k=3\) as the default setting, which requires the deviation to persist for at least four consecutive steps. 
This filters transient noise while avoiding an overly strict criterion that would miss short but safety-relevant deviations in closed-loop driving.

The deviation threshold has a stronger effect. 
As expected, increasing \(\epsilon\) reduces ASR, since larger excess trajectory deviations are harder to induce. 
We use \(\epsilon=0.5\,\mathrm{m}\) as the default closed-loop threshold because it captures meaningful deviation in realized ego motion while still preserving enough sensitivity to detect early closed-loop drift before it compounds into larger failures. 
Stricter thresholds such as \(1.0\)--\(3.0\,\mathrm{m}\) remain useful for measuring more severe deviations, but they undercount smaller perturbation-induced changes that can still become safety-relevant over time.

Table~\ref{tab:semantic_slowdown_k_sensitivity} reports the same temporal-consistency analysis for reasoning objectives. 
Semantic reasoning ASR decreases gradually as \(k\) increases, but remains substantial even under stricter windows, indicating that reasoning shifts are often temporally persistent. 
In contrast, slowdown ASR drops much faster with larger \(k\), especially for Alpamayo1, suggesting that reasoning-length expansion is more intermittent than semantic drift. 
Overall, these results support our default closed-loop choice of \(k=3\): it is strict enough to avoid counting transient fluctuations, but not so strict that it removes meaningful closed-loop deviations.

\begin{table}[t]
\centering
\caption{Closed-loop trajectory ASR sensitivity to temporal consistency \(k\) and deviation threshold \(\epsilon\). ASR requires the excess deviation \(D_a(t)-D_b(t)\) to exceed \(\epsilon\) for at least \(k{+}1\) consecutive steps.}
\label{tab:asr_sensitivity_combined}
\small
\setlength{\tabcolsep}{5pt}
\renewcommand{\arraystretch}{1.15}

\begin{tabular}{lcccccc|ccccc}
\toprule
\multirow{2}{*}{\textbf{Model}} 
& \multicolumn{6}{c|}{\textbf{Temporal Consistency ($\epsilon=0.5$m)}} 
& \multicolumn{5}{c}{\textbf{Deviation Threshold ($k=3$)}} \\

\cmidrule(lr){2-7} \cmidrule(lr){8-12}

& \textbf{$k{=}0$} 
& \textbf{$k{=}1$} 
& \textbf{$k{=}2$} 
& \textbf{$k{=}3$} 
& \textbf{$k{=}5$} 
& \textbf{$k{=}10$} 
& \textbf{$0.5$m} 
& \textbf{$1.0$m} 
& \textbf{$1.5$m} 
& \textbf{$2.0$m} 
& \textbf{$3.0$m} \\

\midrule

Alpamayo1 
& 76.0 & 72.0 & 72.0 & 72.0 & 72.0 & 68.0 
& 72.0 & 58.0 & 50.0 & 44.0 & 32.0 \\

Alpamayo1.5 
& 50.0 & 50.0 & 48.0 & 48.0 & 46.0 & 40.0 
& 48.0 & 34.0 & 30.0 & 26.0 & 20.0 \\

\bottomrule
\end{tabular}
\end{table}

\begin{table*}[t]
\centering
\caption{Closed-loop reasoning ASR sensitivity to temporal consistency \(k\). Semantic deviation uses \(\epsilon=0.5\), while slowdown uses a \(25\%\) excess token-length threshold after subtracting benign variation.}
\label{tab:semantic_slowdown_k_sensitivity}
\small
\setlength{\tabcolsep}{5pt}
\renewcommand{\arraystretch}{1.15}

\begin{tabular}{llccccccc}
\toprule
\multirow{2}{*}{\textbf{Model}} 
& \multirow{2}{*}{\textbf{Metric}}
& \multicolumn{7}{c}{\textbf{ASR across $k$ (\%)}} \\
\cmidrule(lr){3-9}
& 
& \textbf{$k{=}0$}
& \textbf{$k{=}1$}
& \textbf{$k{=}2$}
& \textbf{$k{=}3$}
& \textbf{$k{=}5$}
& \textbf{$k{=}7$}
& \textbf{$k{=}10$} \\
\midrule

\multirow{6}{*}{Alpamayo1}
& Object      & 98.0 & 92.0 & 86.0 & 78.0 & 74.0 & 62.0 & 52.0 \\
& Relation    & 98.0 & 98.0 & 90.0 & 86.0 & 82.0 & 74.0 & 48.0 \\
& Implication & 98.0 & 92.0 & 88.0 & 82.0 & 76.0 & 62.0 & 48.0 \\
& Planning    & 98.0 & 96.0 & 94.0 & 90.0 & 86.0 & 78.0 & 58.0 \\
& Overall     & 98.0 & 92.0 & 88.0 & 86.0 & 74.0 & 60.0 & 50.0 \\
\cmidrule(lr){2-9}
& Slowdown    & 44.0 & 22.0 & 10.0 & 8.0  & 4.0  & 0.0  & 0.0  \\

\midrule

\multirow{6}{*}{Alpamayo1.5}
& Object      & 90.0 & 88.0 & 84.0 & 82.0 & 64.0 & 54.0 & 52.0 \\
& Relation    & 92.0 & 92.0 & 84.0 & 82.0 & 66.0 & 54.0 & 50.0 \\
& Implication & 90.0 & 86.0 & 80.0 & 76.0 & 58.0 & 56.0 & 48.0 \\
& Planning    & 92.0 & 88.0 & 80.0 & 80.0 & 66.0 & 56.0 & 40.0 \\
& Overall     & 90.0 & 86.0 & 78.0 & 78.0 & 64.0 & 56.0 & 50.0 \\
\cmidrule(lr){2-9}
& Slowdown    & 38.0 & 30.0 & 24.0 & 20.0 & 12.0 & 10.0 & 4.0  \\

\bottomrule
\end{tabular}
\end{table*}

\end{document}